\documentclass[aps,prd,twocolumn,twoside,superscriptaddress,showpacs,floatfix]{revtex4}

\usepackage{graphicx}
\usepackage{dcolumn}
\usepackage{bm}
\usepackage{natbib}
\usepackage{multirow}
\usepackage{amsmath}
\usepackage{hyperref}

\newcommand{\apjs}{{Astrophys.~J.~Supp.}}

\newcommand{\mnras}{{Mon.~Not.~R.~Astron.~Soc.}}

\begin{document}

\title{Minimizing the stochasticity of halos in large-scale structure surveys}

\author{Nico Hamaus} \email{hamaus@physik.uzh.ch}
\affiliation{Institute for Theoretical Physics, University of Zurich, 8057 Zurich, Switzerland}
\author{Uro{\v s} Seljak} \email{seljak@physik.uzh.ch}
\affiliation{Institute for Theoretical Physics, University of Zurich, 8057 Zurich, Switzerland}
\affiliation{Physics Department, Astronomy Department and Lawrence Berkeley National Laboratory, University of California, Berkeley, California 94720, USA}
\affiliation{Ewha University, Seoul 120-750, S. Korea}
\author{Vincent Desjacques}
\affiliation{Institute for Theoretical Physics, University of Zurich, 8057 Zurich, Switzerland}
\author{Robert E. Smith}
\affiliation{Institute for Theoretical Physics, University of Zurich, 8057 Zurich, Switzerland}
\author{Tobias Baldauf}
\affiliation{Institute for Theoretical Physics, University of Zurich, 8057 Zurich, Switzerland}

\date{\today}

\begin{abstract}
In recent work (Seljak, Hamaus and Desjacques 2009) it was found that weighting central halo galaxies by halo mass can significantly suppress their stochasticity relative to the dark matter, well below the Poisson model expectation. This is useful for constraining relations between galaxies and the dark matter, such as the galaxy bias, especially in situations where sampling variance errors can be eliminated. In this paper we extend this study with the goal of finding the optimal mass-dependent halo weighting. We use $N$-body simulations to perform a general analysis of halo stochasticity and its dependence on halo mass. We investigate the stochasticity matrix, defined as $C_{ij}\equiv\langle(\delta_i -b_i\delta_m)(\delta_j-b_j\delta_m)\rangle$, where $\delta_m$ is the dark matter overdensity in Fourier space, $\delta_i$ the halo overdensity of the $i$-th halo mass bin, and $b_i$ the corresponding halo bias. In contrast to the Poisson model predictions we detect nonvanishing correlations between different mass bins. We also find the diagonal terms to be sub-Poissonian for the highest-mass halos. The diagonalization of this matrix results in one large and one low eigenvalue, with the remaining eigenvalues close to the Poisson prediction $1/\bar{n}$, where $\bar{n}$ is the mean halo number density. The eigenmode with the lowest eigenvalue contains most of the information and the corresponding eigenvector provides an optimal weighting function to minimize the stochasticity between halos and dark matter. We find this optimal weighting function to match linear mass weighting at high masses, while at the low-mass end the weights approach a constant whose value depends on the low-mass cut in the halo mass function. This weighting further suppresses the stochasticity as compared to the previously explored mass weighting. Finally, we employ the halo model to derive the stochasticity matrix and the scale-dependent bias from an analytical perspective. It is remarkably successful in reproducing our numerical results and predicts that the stochasticity between halos and the dark matter can be reduced further when going to halo masses lower than we can resolve in current simulations.

\end{abstract}


\maketitle

\setcounter{footnote}{0}

\section{Introduction}
The large-scale structure (LSS) of the Universe carries a wealth of information about the physics that governs cosmological evolution. By measuring LSS we can attempt to answer such fundamental questions as what the Universe is made of, what the initial conditions for the structure in the Universe were, and what its future will be. Traditionally, the easiest way to observe it is by measuring galaxy positions and redshifts, which provides the 3D spatial distribution of LSS via so-called redshift surveys (e.g., \cite{SDSS4,SDSS7}).

However, dark matter dominates the evolution and relation to fundamental cosmological parameters, while galaxies are only biased, stochastic tracers of this underlying density field. On large scales, this bias is expected to be a constant offset in clustering amplitude relative to the dark matter, which can be removed to reconstruct the dark matter power spectrum~\cite{Bias}. Nevertheless, this reconstruction is hampered due to a certain degree of randomness in the distribution of galaxies, which is based on the nonlinear and stochastic relation between galaxies and the dark matter. In the simplest model one describes this stochasticity with the Poisson model of \emph{shot noise}. Shot noise constitutes a source of error in the power spectrum \cite{Stochasticity} and therefore limits the accuracy of cosmological constraints. The Poisson model predicts it to be determined by the inverse of the galaxy number density, assuming galaxies to be random and pointlike tracers. However, galaxies are born inside dark matter halos and for these extended, gravitationally interacting objects, the shot noise model is harder to describe. It is thus desirable to develop estimators that are least affected by this source of stochasticity.

In Fourier space the stochasticity of galaxies is usually described by the \emph{cross-correlation coefficient}
\begin{equation}
r_{gm}\equiv \frac{P_{gm}}{\sqrt{\hat{P}_{gg}P_{mm}}}\;,
\end{equation}
where $\hat{P}_{gg}$ is the measured galaxy autopower spectrum, $P_{mm}$ the dark matter autopower spectrum, and $P_{gm}$ the cross-power spectrum of the two components. The cross-correlation coefficient $r_{gm}$ can be related to the shot noise power $\sigma^2$, which is commonly defined via the decomposition $\hat{P}_{gg}=P_{gg}+\sigma^2$, with $P_{gg}=b^2P_{mm}$ and the bias defined as $b=P_{gm}/P_{mm}$. This yields
\begin{equation}
\frac{\sigma^2}{P_{gg}}=\frac{1-r_{gm}^2}{r_{gm}^2} \;. \label{sigma-r}
\end{equation}
Thus, the lower the shot noise, the smaller the stochasticity, i.e., the deviation of the cross-correlation coefficient from unity. Minimizing this stochasticity is important if one attempts to determine the relation between galaxies and the underlying dark matter. One example of such an application is correlating the weak lensing signal, which traces dark matter, to properly radially weighted galaxies \cite{Pen}: an accurate determination of the galaxy bias can be combined with a 3-dimensional galaxy redshift survey to greatly reduce the statistical errors relative to the corresponding 2-dimensional weak lensing survey.

The ultimate precision on how accurate the galaxy bias can be estimated from such methods is determined by the cross-correlation coefficient and previous work has shown that it can deviate significantly from unity for uniformly weighted galaxies or halos \cite{Stochasticity}. However, it was demonstrated recently that weighting halos by mass considerably reduces the stochasticity between halos and the dark matter \cite{sn_letter}. The purpose of this paper is to explore this more systematically and to develop an optimal weighting method that achieves the smallest possible stochasticity.

Our definition of the shot noise above is relevant for the methods that attempt to cancel sampling variance (or \emph{cosmic variance}) \cite{fnl_cv,beta_cv} and this will be our primary motivation in this paper. Alternatively, the shot noise is often associated with its contribution to the error in the power spectrum determination, this error usually being decomposed into sampling variance and shot noise. Sampling variance refers to the fact that in a given volume $V$ the number $N_k$ of observable Fourier modes of a given wave vector amplitude is finite. In the case of a Gaussian random field the relative error in the measured galaxy power spectrum $\hat{P}_{gg}$ due to the sum of the two errors is given by $\sigma_{\hat{P}_{gg}}/\hat{P}_{gg}=1/\sqrt{N_k}$ (each complex Fourier mode has two independent realizations and we only count modes with positive wave vector components). Using the above decomposition of the measured power $\hat{P}_{gg}$ into intrinsic power $P_{gg}$ and shot noise $\sigma^2$, one finds
\begin{equation}
\frac{\sigma_{P_{gg}}}{P_{gg}}=\frac{1}{\sqrt{N_k}}\left(1+\frac{\sigma^2}{P_{gg}}\right) \;. \label{sigma_P}
\end{equation}
This definition is ambiguous, since it leaves the decomposition of the measured power into shot noise and shot noise subtracted power unspecified. Most of the analyses so far have simply assumed the Poisson model, where the shot noise is given by the inverse of the number density of galaxies, $\sigma^2=1/\bar{n}$. A second possibility is to define the shot noise such that the galaxy bias estimator~$\sqrt{P_{gg}/P_{mm}}$ becomes as scale independent as possible. The third way is to define it via the stochasticity between halos and the dark matter, i.e., the cross-correlation coefficient $r_{gm}$ as in Eq.~(\ref{sigma-r}). We choose the third definition, but will comment on the relations to the other two methods as well.

It is important to emphasize here that the first two definitions are not directly related to the applications where the sampling variance error can be eliminated, since they do not include correlations between tracers (where the dark matter itself can also be seen as a tracer).
While in this paper we focus on minimizing the error on the bias estimation using the sampling variance canceling method, there are other applications where correlating dark matter and galaxies, or two differently biased galaxy samples, allows us to reduce the sampling variance error \cite{fnl_cv,beta_cv,Gil-Marin}. In such cases the stochasticity, or the shot noise to power ratio as defined in Eq.~(\ref{sigma-r}), is the dominant source of error and methods capable of reducing it offer the potential to further advance the precision of cosmological tests. Indeed, since the error on the power spectrum as in Eq.~(\ref{sigma_P}) contains two contributions, in the past there was not much interest in investigating the situation where the shot noise is much smaller than sampling variance. It is the situations where the sampling variance error vanishes that are most relevant for our study.

In this paper we will focus on the relation between halos and the underlying dark matter, using two-point correlations in Fourier space (i.e. the power spectrum) as a statistical estimator. A further step to connect halos to observations of galaxies can be accomplished by specification of a halo occupation distribution for galaxies \cite{HOD}, but we do not investigate this in any detail. Alternatively, one can think of the halos as a sample of central halo galaxies from which satellites have been removed.

\section{Shot noise matrix \label{Shot noise}}
The term shot noise is usually related to the fact that the sampling of a continuous field with a finite number of objects yields a spurious contribution of power to its autopower spectrum. In the Poisson model the contribution to the autopower spectrum is $1/\bar{n}$, where $\bar{n}=N/V$ is the mean number density of objects sampling the continuous field, whereas the cross-power spectrum of two distinct samples of objects is not affected (see, e.g., \cite{Peebles,Robert}). However, in cosmology one studies galaxies residing in dark matter halos, which are not a random subsample of the dark matter particles. The Poisson model does not account for that fact.

In recent work it has been argued that there are other nonlinear terms that appear like white noise terms in the power spectrum of halos and so a more general approach is needed to determine the shot noise \cite{McDonald}. In order to account for this fact we define the shot noise more generally as the two-point correlation matrix
\begin{equation}
C_{ij}\equiv\langle(\delta_i-b_i\delta_m)(\delta_j-b_j\delta_m)\rangle \;. \label{sn}
\end{equation}
Here the subscripts $i$ and $j$ refer to specific subsamples of the halo density field with overdensities $\delta_i$ and $\delta_j$ and corresponding scale-independent bias $b_i$ and $b_j$, respectively. The dark matter density fluctuation is denoted by $\delta_m$ and the angled brackets denote an ensemble average. We work in Fourier space and the $\delta$'s are the complex Fourier components of the density field,
\begin{equation}
\delta(\mathbf{k})=\frac{1}{\sqrt{V}}\int\delta(\mathbf{x})e^{-i\mathbf{k\cdot x}}\mathrm{d}^3x \;.
\end{equation}
However, we handle the complex density modes $\delta$ as real quantities, since their real and imaginary parts are uncorrelated and one can treat them as two independent modes. Further, we assume the overdensity of a particular halo sample $i$ to be composed of two terms~\cite{Stochasticity}:
\begin{equation}
\delta_i=b_i\delta_m+\epsilon_i\;,\label{delta_i}
\end{equation}
where $\epsilon_i$ is a random variable of zero mean assumed to be uncorrelated with the signal, i.e., $\langle\epsilon_i\delta_m\rangle=0$. It follows that the bias parameter $b_i$ can be obtained from cross correlation with the dark matter,
\begin{equation}
b_i=\frac{\langle\delta_i\delta_m\rangle}{\langle\delta_m^2\rangle} \;, \label{bias}
\end{equation}
and the shot noise matrix can be written as $C_{ij}=\langle\epsilon_i\epsilon_j\rangle$. With these definitions the cross-correlation coefficient between any given halo bin $i$ and the dark matter,
\begin{equation}
r_{im}\equiv\frac{\langle\delta_i\delta_m\rangle}{\sqrt{\langle\delta_i^2\rangle\langle\delta_m^2\rangle}} \;, \label{cross}
\end{equation}
becomes unity when we subtract the shot noise component $C_{ii}$ from $\langle\delta_i^2\rangle$. We thus recover the shot noise definition from Eq.~(\ref{sigma-r}) and define $P_{ii}\equiv\langle\delta_i^2\rangle-C_{ii}$ to be the halo autopower spectrum (shot noise subtracted), $P_{im}=\langle\delta_i\delta_m\rangle$ the halo-matter cross-power spectrum, and $P_{mm}=\langle\delta_m^2\rangle$ the matter autopower spectrum (we assume the shot noise of the matter auto-, as well as the halo-matter cross-power spectrum to vanish). Note that these relations are still self-consistent if we allow the bias $b_i$ to be scale dependent. Here we will, however, explore the simpler case assuming scale-independent bias, which is a good approximation on large scales.

In the Poisson model the shot noise matrix $C_{ij}$ is diagonal, but this is not necessarily the case in our definition. The objective of this paper is to study all the components of this matrix using $N$-body simulations. In particular we divide the halos into bins of different mass, but equal number density. A diagonalization of the shot noise matrix will then provide its eigenvalues and eigenvectors, which contain important information about the stochastic properties of the halo density field.

\section{Simulations \label{simulations}}
We use the \scshape zHorizon \rm simulations \cite{Robert}, 30 realizations of numerical $N$-body simulations with $750^3$ particles of mass $5.55\times10^{11}h^{-1}\mathrm{M}_{\odot}$ and a box-size of $1.5h^{-1}\mathrm{Gpc}$ (total effective volume of $V_{\mathrm{tot}}=101.25h^{-3}\mathrm{Gpc}^3$) to accurately sample the density field of cold dark matter. The simulations were performed at the University of Zurich supercomputers \scshape zbox2 \rm and \scshape zbox3 \rm with the \scshape gadget ii \rm code \cite{Gadget}. We chose the cosmological parameters to be close to the outcome of the WMAP5 data release \cite{WMAP5}, namely $\Omega_m=0.25$, $\Omega_{\Lambda}=0.75$, $\Omega_b=0.04$, $\sigma_8=0.8$, $n_s=1.0$ and $h=0.7$. The transfer function was computed with the \scshape cmbfast \rm code \cite{CMBFAST} and the initial conditions were set up at redshift $z=50$ with the \scshape 2lpt \rm initial conditions generator \cite{2LPT,2LPT2}.

We applied the friends-of-friends (FoF) algorithm {\scshape b-fof} by V. Springel with a linking length of $20\%$ of the mean interparticle distance and a minimum of 30 particles per halo to generate halo catalogs. The resulting catalogs contain about $1.3\times10^6$ halos ($\bar{n}\simeq3.7\times10^{-4}h^3\mathrm{Mpc^{-3}}$) with masses between $M_{\mathrm{min}}\simeq1.1\times10^{13}h^{-1}\mathrm{M}_{\odot}$ and $M_{\mathrm{max}}\simeq3.1\times10^{15}h^{-1}\mathrm{M}_{\odot}$. In order to investigate the influence of the mass resolution on our results, we employ another set of 5 $N$-body simulations \cite{Vincent} of box-size $1.6h^{-1}\mathrm{Gpc}$ with $1024^3$ particles of mass $3.0\times10^{11}h^{-1}\mathrm{M}_{\odot}$, resolving halos down to $M_{\mathrm{min}}\simeq5.9\times10^{12}h^{-1}\mathrm{M}_{\odot}$ ($\bar{n}\simeq7.0\times10^{-4}h^3\mathrm{Mpc^{-3}}$). All other parameters of this simulation are similar to the one above, namely $\Omega_m=0.279$, $\Omega_{\Lambda}=0.721$, $\Omega_b=0.046$, $\sigma_8=0.81$, $n_s=0.96$, $h=0.7$. One further realization with these parameters was generated with an even higher mass resolution, namely $1536^3$ particles of mass $4.7\times10^{10}h^{-1}\mathrm{M}_{\odot}$ in a box of $1.3h^{-1}\mathrm{Gpc}$, resolving halos down to $M_{\mathrm{min}}\simeq9.4\times10^{11}h^{-1}\mathrm{M}_{\odot}$ ($\bar{n}\simeq4.0\times10^{-3}h^3\mathrm{Mpc^{-3}}$).

The density fields of dark matter and halos in configuration space were computed via interpolation of the particles onto a cubical mesh with $512^3$ grid points using a cloud-in-cell mesh assignment algorithm \cite{CIC}. We then applied fast Fourier transforms to compute the modes of the fields in $k$-space. All our results are presented at $z=0$ and we do not explore the redshift dependence, because at higher redshifts the halo number density is lower and we wish to explore the stochastic properties of halos in the high density limit.

\section{Analysis \label{analysis}}

\subsection{Estimators for the binned halo density field}
\begin{figure}[!t]
\centering
\resizebox{\hsize}{!}{\includegraphics[viewport=1cm 2.1cm 19cm 13.1cm,clip]{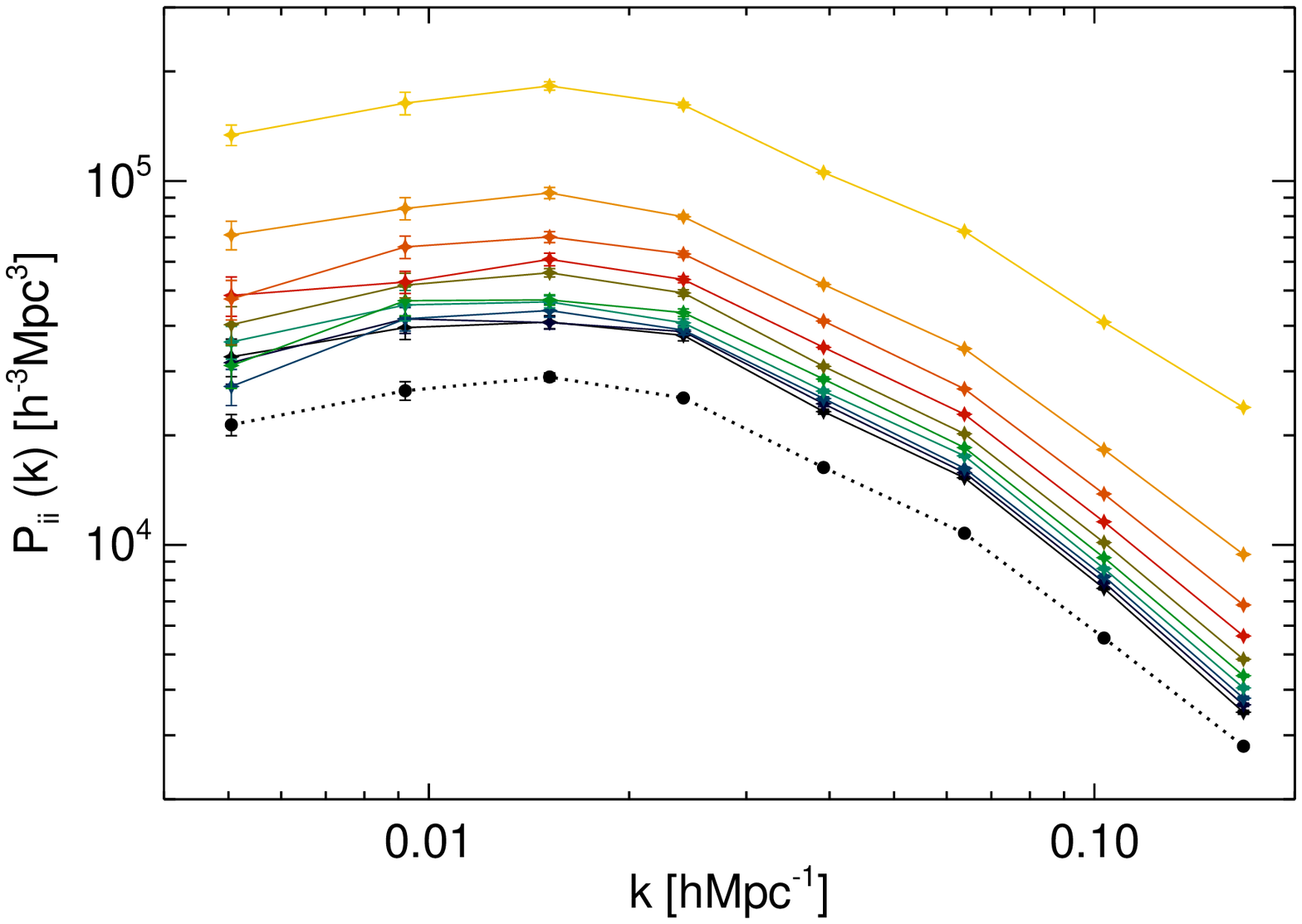}}
\resizebox{\hsize}{!}{\includegraphics[viewport=1cm 2.1cm 19cm 13.1cm,clip]{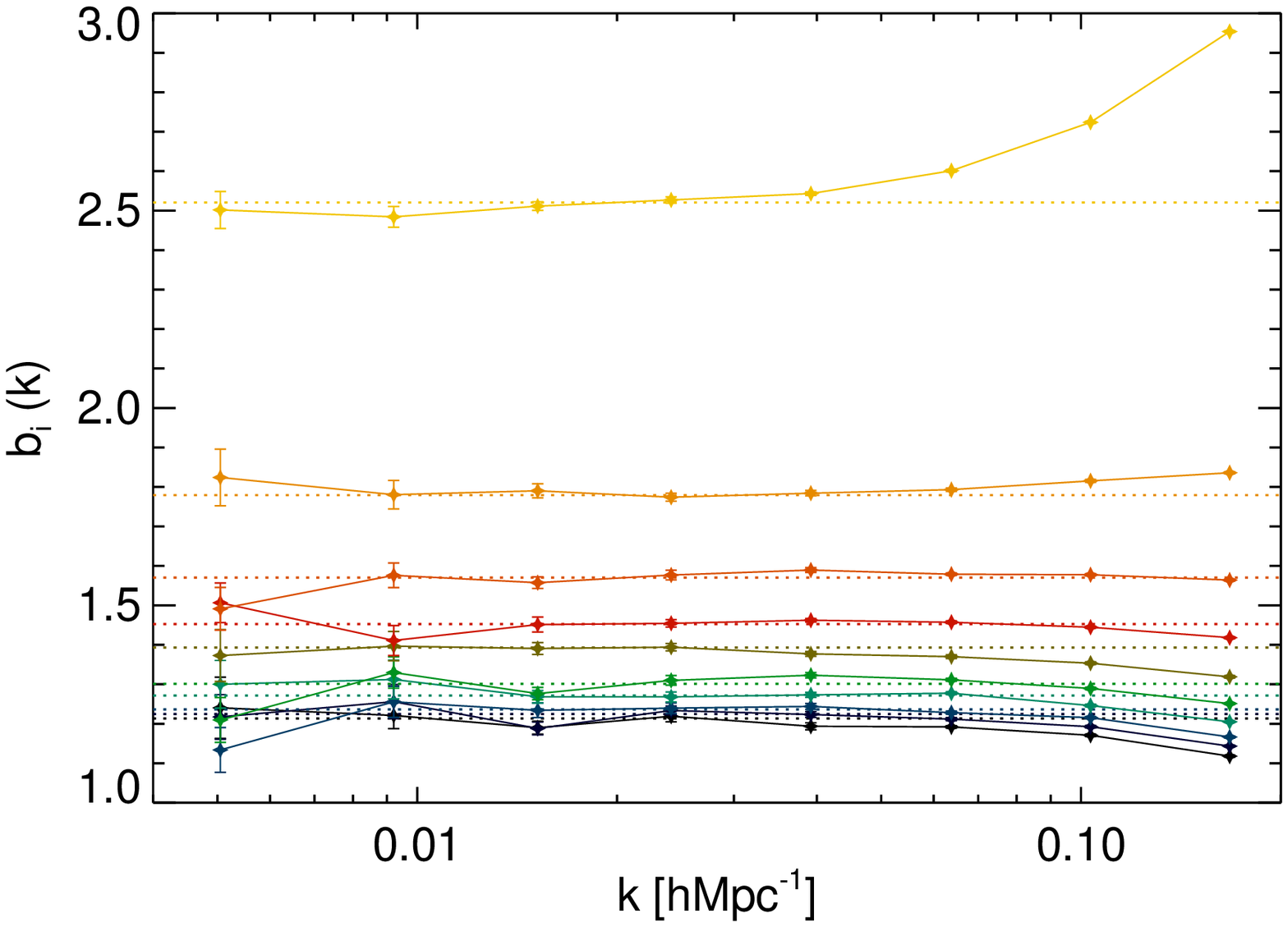}}
\resizebox{\hsize}{!}{\includegraphics[viewport=1cm 0.4cm 19cm 13.1cm,clip]{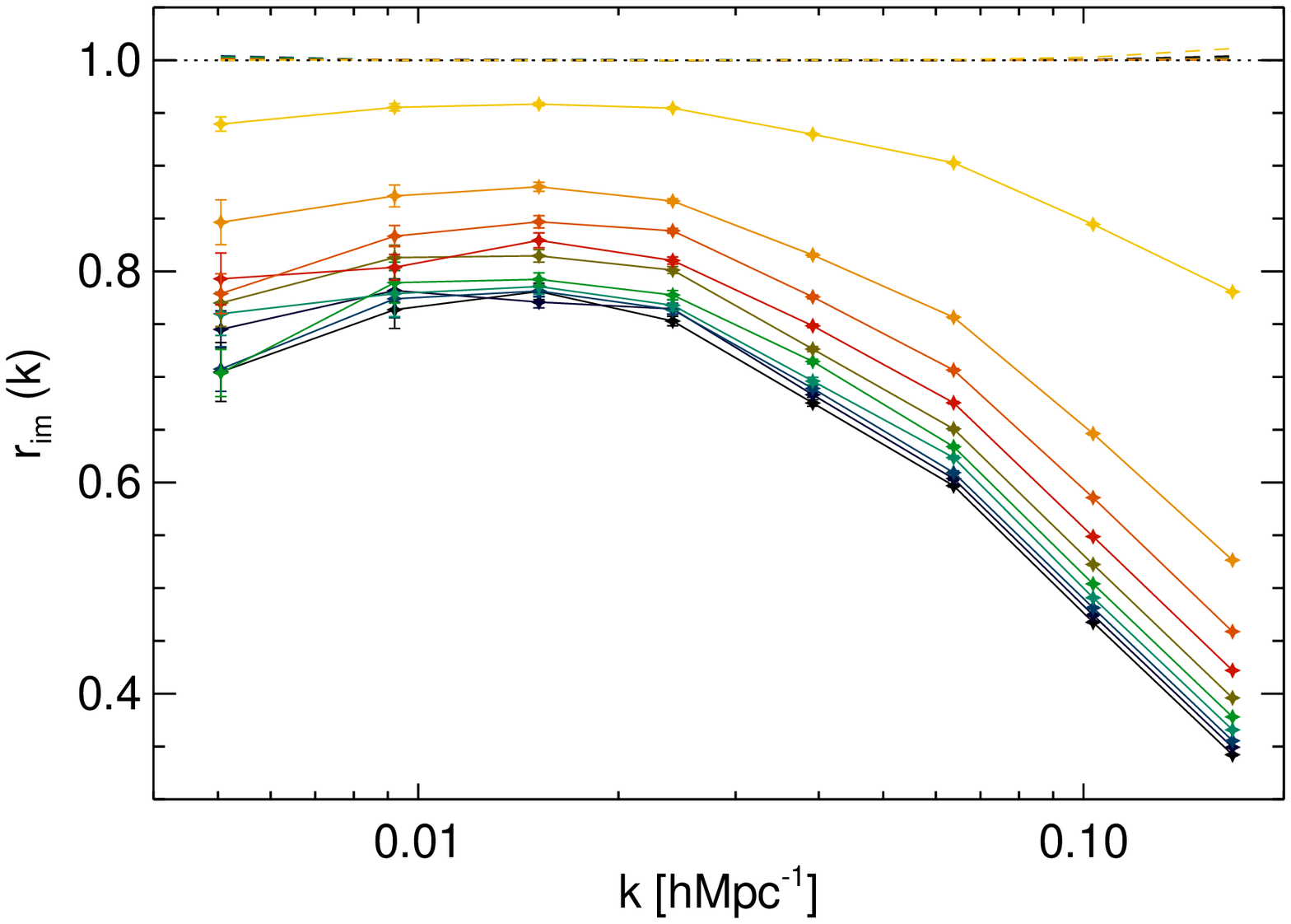}}
\caption{TOP: Autopower spectra for 10 consecutive halo mass bins (solid colored lines) and the dark matter (dotted black line). MIDDLE: Bias of the 10 halo bins determined from the cross power with the dark matter, the dotted lines show the scale-independent bias. BOTTOM: Cross-correlation coefficients of the 10 halo bins with the dark matter (solid colored lines) without shot noise subtraction. When the shot noise $C_{ii}$ is subtracted from $\langle\delta_i^2\rangle$, by definition the cross-correlation coefficient becomes unity (dashed lines). For reference, the value $r=1$ is plotted (dotted black line). The error bars on all three plots were computed from the ensemble of the 30 independent realizations of our simulations. They show the standard deviation on the mean of each quantity shown.}
\label{10bins}
\end{figure}

The shot noise matrix from Eq.~(\ref{sn}) is calculated by plugging in the Fourier modes provided by our simulations and averaging over a range of wave numbers. The bias is determined via the ratio in Eq.~(\ref{bias}), we thus neglect any shot noise contribution in this expression. In Eq.~(\ref{sn}) we use the scale-independent bias, which is obtained by averaging over the range $k\le0.024\;h\mathrm{Mpc}^{-1}$, corresponding to our first four $k$-bins. This range of wave numbers is least affected by scale dependence, as apparent from the middle panel of Fig.~\ref{10bins}.

For the division into subsamples we bin the full halo catalog into bins of different mass, keeping the number density of each bin constant. This is done by sorting the halos according to their mass and then dividing this sorted array into subarrays with an equal number of halos. We use 10 bins for most of the plots presented here, since more bins make them increasingly hard to read. For some plots we also show the results with 30 and 100 bins to provide a more accurate sampling of halo masses. A convergence of the results can only be reached with infinitely many bins, which is numerically impossible to accomplish. However, using linear mass weighting of the halos within each bin makes the results converge faster, as will be justified later. We apply this technique to our 100 halo mass bins, as shown in some of the following plots.

\subsubsection{Power spectrum, bias and cross-correlation coefficient}
We start by looking at the autopower spectrum of the halos in each mass bin as shown in the top panel of Fig.~\ref{10bins}, using our lower resolution simulation with average halo number density of $\bar{n}\simeq3.7\times10^{-4}h^3\mathrm{Mpc^{-3}}$. The halo autopower spectra have been subtracted by $C_{ii}$, the diagonal elements of the shot noise matrix from Eq.~(\ref{sn}), depicted below in Fig.~\ref{snfig}. The halo subsamples increasingly gain power with higher mass due to their enhanced bias, which is plotted in the middle panel of Fig.~\ref{10bins}. This plot shows the bias obtained from Eq.~(\ref{bias}) as a function of $k$. The scale-independent bias is drawn as straight dotted lines for comparison. On large scales, roughly below $k\simeq0.015\;h\mathrm{Mpc}^{-1}$, sampling variance makes the curves appear more noisy, while on smaller scales, $k\gtrsim0.04\;h\mathrm{Mpc}^{-1}$, possibly nonlinear evolution of the density field or higher-order bias corrections set in causing the halo bias to pick up a scale dependence \cite{Robert-2007}. This scale dependence is most pronounced for the highest-mass bin.

The degree of halo stochasticity can also be assessed in the cross-correlation coefficient between halos and the dark matter, as depicted in the bottom panel of Fig.~\ref{10bins}. We see that the more massive halos are a \emph{less stochastic} tracer of the dark matter. Note that subtracting our definition of the shot noise from the autocorrelation of halos makes the cross-correlation coefficient become unity. It has the nice property that the bias determined from halo auto-correlation and from halo-matter cross-correlation is identical by definition.

\subsubsection{Shot noise matrix}
\begin{figure}[!t]
\centering
\resizebox{\hsize}{!}{\includegraphics[viewport=0.3cm 0.4cm 19cm 13.1cm,clip]{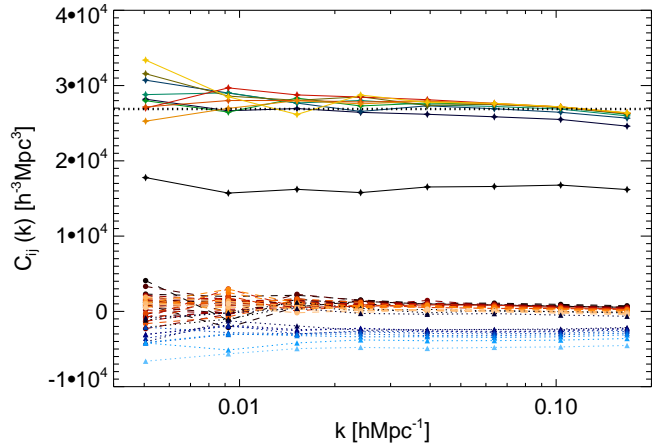}}
\caption{Elements of the shot noise matrix as defined in Eq.~(\ref{sn}) with 10 halo mass bins. Most of the diagonal components (solid lines with stars) agree with Poisson white noise, i.e., $C_{ii}=1/\bar{n}_i$ (dotted black line on top), except the highest-mass bin which is clearly suppressed (solid black line). There are both positive (dashed lines with circles, scaled in red) and negative (dotted lines with triangles, scaled in blue) off-diagonal components.}
\label{snfig}
\end{figure}

We now turn to the calculation of the shot noise matrix for the 10 halo mass bins. Figure~\ref{snfig} shows each element of $C_{ij}$ plotted against the wave number. As expected, the diagonal components of the shot noise (solid lines) are dominant. They all show essentially no scale dependence and match the usual expectation of $1/\bar{n}_i$ very well (where $\bar{n}_i$ is the mean number density of halos in bin~$i$), except for the highest-mass bin (solid, black line), which is suppressed by about a factor of 2. The conventional expression for the shot noise breaks down for the highest-mass halos. This sub-Poissonian behavior of the shot noise at high masses has been found in simulations before \cite{Casas,Manera}.

Moreover we find both negative and positive elements in the off-diagonal parts of the shot noise matrix. In the case of $N$ halo bins, in total there are $N(N+1)/2$ independent elements, since $C_{ij}$ is a symmetric matrix. These are composed of $N$ diagonal and $N(N-1)/2$ off-diagonal elements. Hence, in the case of $N=10$, there are 45 off-diagonal elements and we find 33 of them to be positive (dashed lines), while 12 are negative (dotted lines). While all off-diagonal elements are white noise like, i.e., scale independent, most of the negative components have a higher magnitude than the positive ones. The former correspond to the cross correlations of any given halo mass bin with the highest-mass halos.

This finding is rather surprising, because shot noise cross correlations are usually being neglected. Since there appear to be negative off-diagonal components in the shot noise matrix and their magnitude exceeds the positive ones, one might expect that a suitable linear combination of the halo bins can reduce the total shot noise, as found in \cite{sn_letter}. In the subsequent section we will show that this expectation is indeed fulfilled.

\subsection{Eigensystem of the shot noise matrix}
In order to find the principal components of the shot noise matrix we have to diagonalize it by determining its eigenvalues $\lambda^{(l)}$ and eigenvectors $V^{(l)}$, defined via
\begin{equation}
\sum_jC_{ij} V^{(l)}_j=\lambda^{(l)} V^{(l)}_i \;.
\end{equation}
The superscript $(l)$ is used to enumerate the eigenvalues and eigenvectors, while the subscripts $i$ and $j$ refer to the components of the vectors and matrices. We use routines from \cite{NR} to do the calculations.

\begin{figure*}[!t]
\centering
\resizebox{\hsize}{!}{
\includegraphics[viewport=0.6cm 0.4cm 19cm 13.2cm,clip]{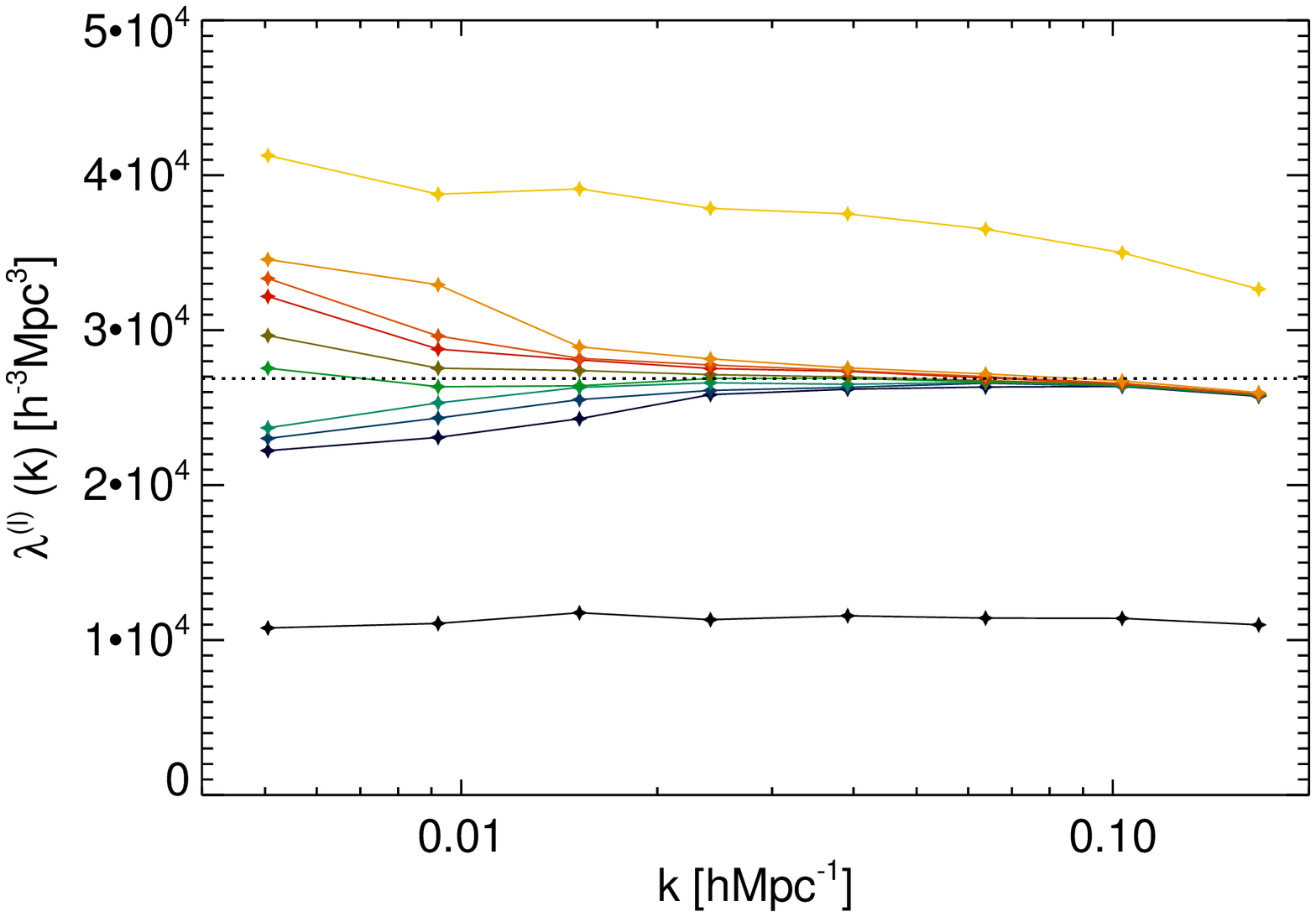}
\qquad
\includegraphics[viewport=1cm 0.4cm 19cm 13.1cm,clip]{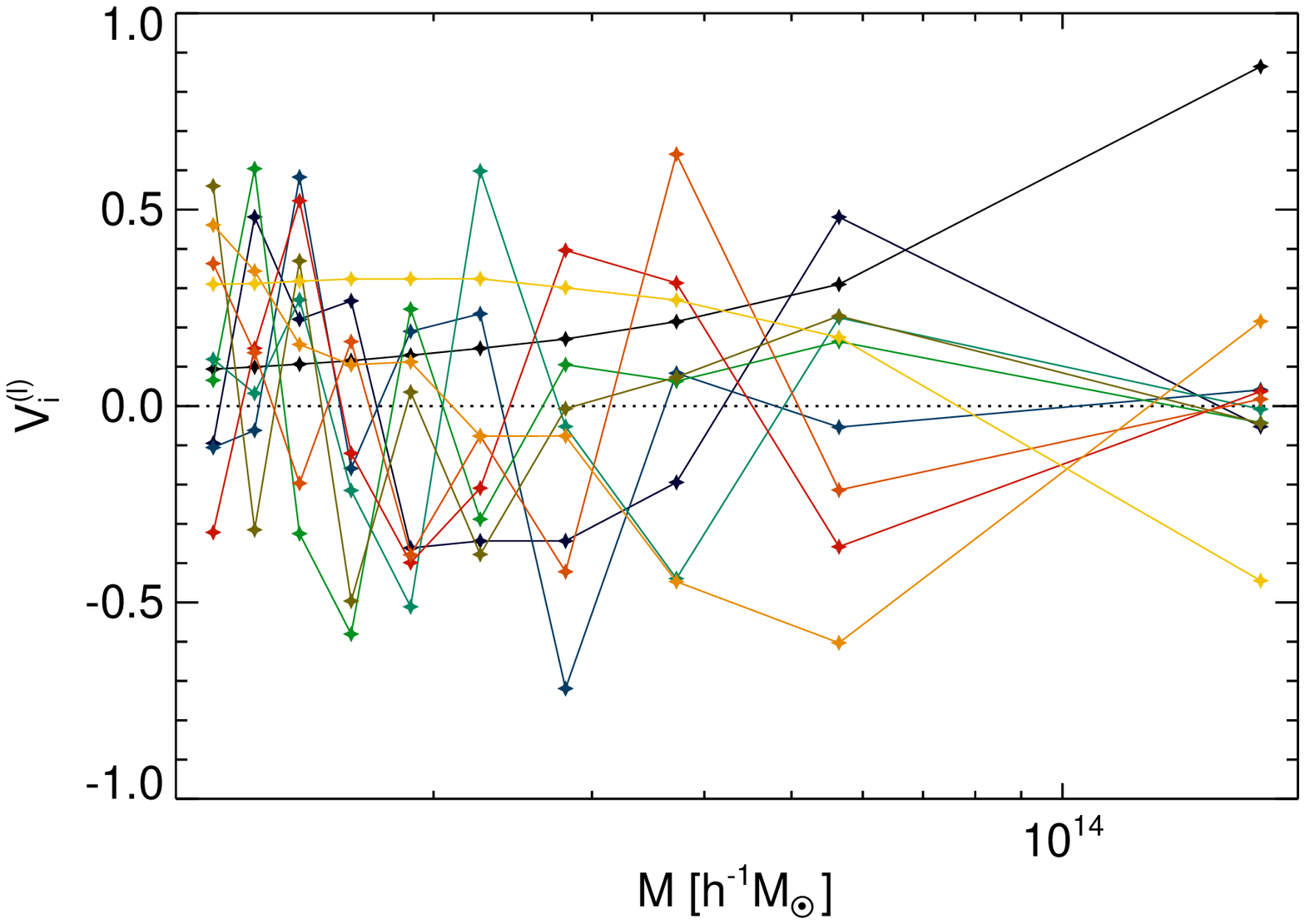}}
\caption{The 10 eigenvalues (left) and eigenvectors (right) of the shot noise matrix from Fig.~\ref{snfig} in corresponding colors. The black dotted line shows the value $1/\bar{n}_i$. The eigenvectors are averaged over the entire $k$-range.}
\label{sn_ev}
\end{figure*}

\subsubsection{Eigenvalues}
The left panel of Fig.~\ref{sn_ev} shows the eigenvalues $\lambda^{(l)}$ of the shot noise matrix from Fig.~\ref{snfig} for the 10 halo bins as a function of $k$. The eigenvalues are computed separately for every $k$-bin and then ordered by their magnitude. As apparent from the figure, we find two eigenvalues to differ significantly from all the others. One of them is enhanced by roughly a factor of 1.5 and one is suppressed by a factor of about 2.5 compared to the other ones that lie close to the value $1/\bar{n}_i$. The spread of the curves at low $k$ is likely due to the low number of modes available there, making the eigenvalue determination inaccurate.

This result reveals the fact that one of the eigenvectors, which represents a particular linear combination of the halo mass bins, yields a very low shot noise level. This shot noise level is determined by the lowest eigenvalue of the shot noise matrix, which we will denote as $\lambda^{-}$. Increasing the number of halo bins we find an even stronger suppression of $\lambda^{-}$ compared to the expectation of $1/\bar{n}_i$ (see Sec.~\ref{Signal-to-Noise}).

The other eigenvalue that differs from the value $1/\bar{n}_i$ represents the highest shot noise level. We designate this eigenvalue $\lambda^{+}$. Since it does not carry much information (see below) we do not investigate it further in this paper beyond noticing that it is likely to be connected to the second-order bias. We note that had we investigated the halo covariance matrix $\langle\delta_i\delta_j\rangle$, we  would not have been able to reveal the lowest eigenvalue as cleanly, because it would have been swamped by sampling variance. Indeed, previous work focused its attention mostly on the largest eigenvalues of $\langle\delta_i\delta_j\rangle$ \cite{Bonoli}.

\subsubsection{Eigenvectors}
Every eigenvector $V_i^{(l)}$ is a function of the wave number, just like the eigenvalues. However, as can be seen from Figs.~\ref{snfig} and \ref{sn_ev}, over a reasonable range of wave numbers this dependence can be ignored and we average the eigenvectors over the entire $k$-range. We also divide each eigenvector by its length $|V_i|=(\sum_iV_i^2)^{1/2}$ to normalize it.

The right panel of Fig.~\ref{sn_ev} displays the $10$ eigenvectors corresponding to the $10$ eigenvalues in the left panel. Each component of an eigenvector corresponds to one halo mass bin. Since we have equal number densities per bin, the mass range per bin gets wider with increasing mass due to the rapid decline of the halo mass function. Every data point in the figure is plotted at the respective average mass of each halo bin. Only one eigenvector shows exclusively positive components, while at least one negative component can be found in the remaining eigenvectors. It is this eigenvector that corresponds to the lowest eigenvalue $\lambda^{-}$ and we will denote it as $V_i^{-}$. Its components continuously increase with mass. The eigenvector $V_i^{+}$ corresponding to the largest eigenvalue $\lambda^{+}$ also shows a monotonic behavior, but its components decrease with mass and turn negative at the high-mass end (similarly to the second-order halo bias with an opposite sign).

One can think of the eigenvectors as weighting functions for the halos, since each component acts as a weight for the associated halo mass bin. Hence, the weighted halo density field $\delta_w(\mathbf{x})$ in configuration space can be written as a weighted sum over the halo mass bins, normalized by the sum of the weights,
\begin{equation}
\delta_w^{(l)}(\mathbf{x})=\frac{\sum_i V^{(l)}_i\delta_i(\mathbf{x})}{\sum_i V^{(l)}_i} \;. \label{delta_w}
\end{equation}

\begin{figure}[!t]
\centering
\resizebox{\hsize}{!}{\includegraphics[scale=0.45,viewport=1cm 0.4cm 19cm 13.1cm,clip]{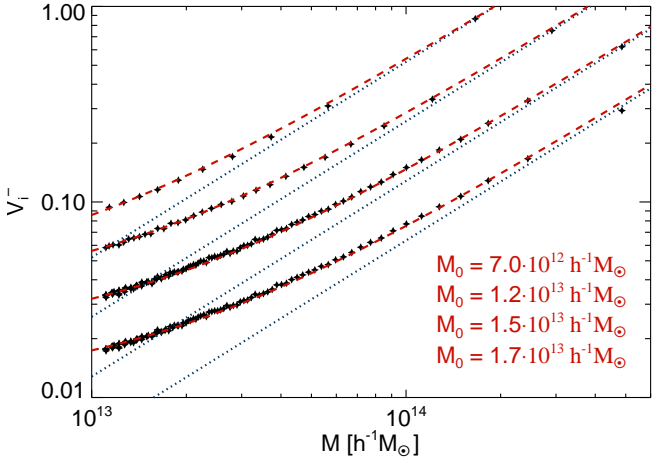}}
\caption{The normalized eigenvector $V_i^{-}$, corresponding to the lowest shot noise eigenvalue $\lambda^{-}$, computed for 10, 30, 100 uniformly weighted bins and 100 mass-weighted bins (from top to bottom). The latter was shifted downwards by a factor of 2 for visibility. The dotted (blue) and the dashed (red) lines represent linear and modified mass weighting, respectively. The best-fit values for $M_0$ are given in the bottom right for the respective cases.}
\label{sn_evec1}
\end{figure}

We specifically want to investigate the eigenvector $V_i^{-}$, since it yields the lowest eigenvalue of the shot noise matrix, $\lambda^{-}$. In Fig.~\ref{sn_evec1} we plot the components of $V_i^{-}$ in a log-log plot to investigate this eigenvector in more detail and compare to the results with 30 and 100 mass bins. The components of $V_i^{-}$ increase linearly with mass above $M\simeq10^{14}h^{-1}\mathrm{M}_{\odot}$, while at lower masses the slope tends to become shallower. We compare this eigenvector to two different smooth weighting functions for the halo density field. The first weighting function simply takes the halo mass as a weight for each halo, $w(M)=M$, we will denote it as \emph{linear mass weighting}. However, as apparent from the dotted lines in Fig.~\ref{sn_evec1}, it only matches the components of $V_i^{-}$ at high mass. In order to account for the saturation effect at low masses, we consider a second weighting function that mimics this behavior,
\begin{equation}
w(M)=M+M_0 \;. \label{mod_weight}
\end{equation}
The free parameter $M_0$ determines the shape of this weighting function, it specifies the mass threshold where the saturation sets in. For $M\ll M_0$, Eq.~(\ref{mod_weight}) approaches uniform weighting, whereas in the limit $M\gg M_0$ it matches linear mass weighting. We call this weighting scheme \emph{modified mass weighting}, it is shown as a dashed curve in Fig.~\ref{sn_evec1} and obviously provides a much better fit to $V_i^{-}$ than linear mass weighting. The fit is shown for each case of our mass binning. The best-fit value for $M_0$ increases with the number of bins and in the case of 100 mass-weighted bins becomes $M_0\simeq1.7\times10^{13}h^{-1}\mathrm{M}_{\odot}$. Note that for visibility reasons this eigenvector is shifted downwards by a factor of 2 in the plot.

Similar weighting schemes have already been applied to the halo density field in \cite{sn_letter}, where a significant reduction of the stochasticity between halos and the dark matter could be achieved. In particular, a trial weighting function also denoted as modified mass weighting was shown to improve on linear mass weighting. However, in that paper the functional form was found empirically and was not demonstrated to be optimal. In this work we show why modified mass weighting as defined in Eq.~(\ref{mod_weight}) is the optimal weighting to suppress the stochasticity in halos: in the limit of many halo mass bins it converges to the components of $V_i^{-}$, the eigenvector of the shot noise matrix with the lowest eigenvalue. What remains to be shown is what determines the value of $M_0$: we will argue it depends on the lower boundary of the halo mass function considered.

These results also justify why applying linear mass weighting to the halos \emph{within} each bin leads to a better convergence of the eigenvector towards a smooth weighting with infinitely many bins: linear mass weighting already reduces the stochasticity of each bin as compared to the uniformly weighted case (see \cite{sn_letter}). The resulting eigenvector is then determined more accurately, corresponding to an effectively higher sampling with more bins. This mainly has an effect on the highest-mass bins, since they have the broadest range in mass. At low masses the bins are increasingly narrow, so uniform and linear mass weighting become increasingly similar within a given bin.

Other attempts to find an optimal weighting scheme for the halo (galaxy) density field found the halo bias to yield the best constraining power on dark matter statistics and cosmological parameters when used as weighting function \cite{Bonoli,Percival,Slosar}. We have tried using only $b(M)$ as a weighting function, but found less suppression in halo stochasticity as compared to modified mass weighting.

\subsection{Signal-to-noise \label{Signal-to-Noise}}
While we have demonstrated that it is possible to suppress the stochasticity of a given halo density field using a single linear combination of halo bins, it remains to be shown how the information content in this single eigenmode compares to the complete information content. We cannot answer this question in general, since it depends not only on the property of the shot noise matrix, but also on derivatives of the halo density field with respect to the cosmological parameters one wants to estimate. Those two ingredients depend on halo mass and determine the Fisher information content of the halo density field. We will not explore the general case here and instead focus on the simple case where the information content is expressed via the ratio of the autopower spectrum to the shot noise of a particular tracer (signal-to-noise ratio per mode). Its inverse appears in Eqs.~(\ref{sigma-r}) and (\ref{sigma_P}). We compute it for the weighted halo density field,
\begin{equation}
\frac{S}{N}\equiv\frac{P_w}{\sigma_w^2}=\frac{b_w^2}{\sigma_w^2}P_{mm}=\frac{r_{wm}^2}{1-r_{wm}^2} \;. \label{signal-to-noise}
\end{equation}
$P_w$ denotes the autopower spectrum of the weighted halo density field, $\sigma_w^2$ its shot noise, $b_w$ the corresponding weighted bias and $r_{wm}$ the cross-correlation coefficient between the fields $\delta_w$ and $\delta_m$ as defined in Eq.~(\ref{cross}). The weighted bias can be computed from the halo bins via
\begin{equation}
b_w=\frac{\langle\delta_w\delta_m\rangle}{\langle\delta_m^2\rangle}=\frac{\sum_iV_ib_i}{\sum_iV_i} \;.
\end{equation}
It is clear from this expression that in order to maximize the signal, the eigenvector components should all be of equal sign, since otherwise different halo bins cancel each others signal. Using Eq.~(\ref{sn}) to express $\sigma_w^2$ in terms of the weighted density field $\delta_w$ and Eq.~(\ref{delta_w}) for the definition of $\delta_w$ (we omit the superscripts for clarity), we have
\begin{eqnarray}
\sigma_w^2&\equiv&\langle(\delta_w-b_w\delta_m)^2\rangle=\langle\left(\frac{\sum_iV_i\delta_i}{\sum_iV_i}-\frac{\sum_iV_ib_i}{\sum_iV_i}\delta_m\right)^2\rangle= \nonumber \\
&=&\frac{\sum_{i,j}V_iV_jC_{ij}}{\sum_{i,j}V_iV_j}=\lambda\;\frac{\sum_iV_i^2}{\left(\sum_iV_i\right)^2}\; , \label{sigma_w}
\end{eqnarray}
where $\sum_iV_i^2=1$ in case the eigenvectors are normalized. Hence, the signal-to-noise ratio becomes
\begin{equation}
\frac{S}{N}=\frac{\left(\sum_iV_ib_i\right)^2}{\lambda\sum_iV_i^2}P_{mm}=\frac{b_w^2}{\lambda}\frac{\left(\sum_iV_i\right)^2}{\sum_iV_i^2}P_{mm} \;. \label{s_n}
\end{equation}
Note that this is only the signal-to-noise ratio for one particular weighting of the halo density field, corresponding to one eigenmode of the shot noise matrix. The complete information content of the halos is calculated by summing up all $N$ contributions,
\begin{equation}
\frac{S}{N}=\sum_{l=1}^N\left(\frac{b_w^{(l)}}{\sigma^{(l)}_w}\right)^2P_{mm} \;. \label{sum_s_n}
\end{equation}
However, since we find one very low eigenvalue of the shot noise, most of the signal will be contained in the halos weighted by~$V_i^{-}$. Adding up the denominators of Eq.~(\ref{sum_s_n}) and taking the inverse yields the total noise contribution of the halos. We call it the \emph{reduced} shot noise,
\begin{equation}
\sigma_r^2=\frac{1}{\sum_l(1/\sigma_w^{(l)})^2} \;. \label{sigma_r}
\end{equation}

\begin{figure}[!t]
\centering
\resizebox{\hsize}{!}{\includegraphics[scale=0.45,viewport=0.9cm 0.4cm 19cm 13.1cm,clip]{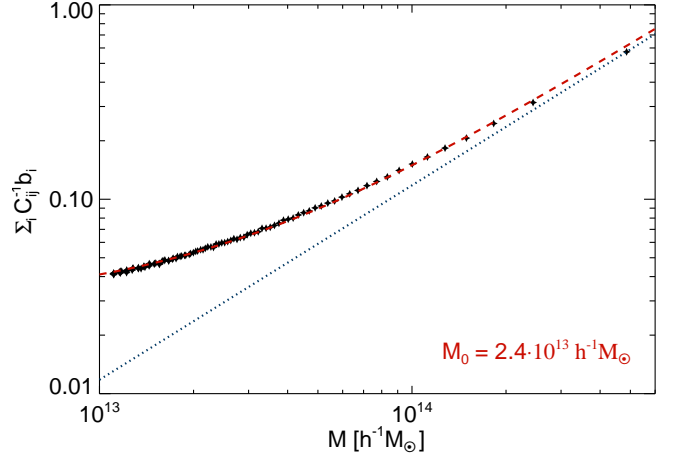}}
\caption{The normalized vector $\sum_{i}C_{ij}^{-1}b_i$ that provides an optimal estimator for the dark matter, computed for 100 uniformly weighted mass bins. Since this vector is very similar to $V_i^{-}$, modified mass weighting (dashed red line) still yields a reasonable fit with a slightly higher value of $M_0$ (bottom right).}
\label{opt_vector}
\end{figure}

Alternatively, the signal-to-noise ratio of the halo density field can be derived from a $\chi^2$ distribution. Since the modes of the halo bins are assumed to be independent, normally distributed variables, the expression
\begin{equation}
\chi^2\equiv\sum_{i,j,k}\left(\delta_i-b_i\delta_m\right)C_{ij}^{-1}\left(\delta_j-b_j\delta_m\right)
\end{equation}
follows a $\chi^2$ distribution. Here, $C_{ij}^{-1}$ refers to the $i$-$j$ component of the inverse shot noise matrix and the index $k$ connotes a summation over all Fourier modes. The derivative of the $\chi^2$ distribution with respect to the inferred dark matter density field $\delta_m$ must vanish, $\partial\chi^2/\partial\delta_m=0$. This yields
\begin{equation}
\delta_m=\frac{\sum_{i,j,k}C_{ij}^{-1}b_i\delta_j}{\sum_{i,j,k}C_{ij}^{-1}b_ib_j} \;. \label{mass_estimator}
\end{equation}
Here, the vector $\sum_{i}C_{ij}^{-1}b_i$ conducts a weighting of the halo mass bins again. The difference to the weighting with one particular eigenvector of $C_{ij}$ is that this vector contains the complete information of all eigenmodes and thus provides an optimal estimator for the dark matter density field. However, as can be seen in Fig.~\ref{opt_vector}, it has a very similar shape as $V_i^{-}$ and modified mass weighting provides an equally thorough fit to this vector. The only difference is a slight increase in the best-fit value for the parameter $M_0$.

The second derivative of the $\chi^2$ distribution leads to the signal-to-noise ratio of the halo density field,
\begin{equation}
\frac{S}{N}\equiv\frac{S}{2N_k}\frac{\partial^2\chi^2}{\partial\delta_m^2}=\sum_{i,j}C_{ij}^{-1}b_ib_jP_{mm} \;, \label{S_N_chi^2}
\end{equation}
This expression is equivalent to the Fisher information on the dark matter density mode $\delta_m$. Here, the reduced shot noise is simply computed as
\begin{equation}
\sigma_r^2=\frac{1}{\sum_{i,j}C_{ij}^{-1}} \;.
\end{equation}

\begin{figure*}[!t]
\centering
\resizebox{\hsize}{!}{
\includegraphics[viewport=1cm 0.4cm 19cm 13.1cm,clip]{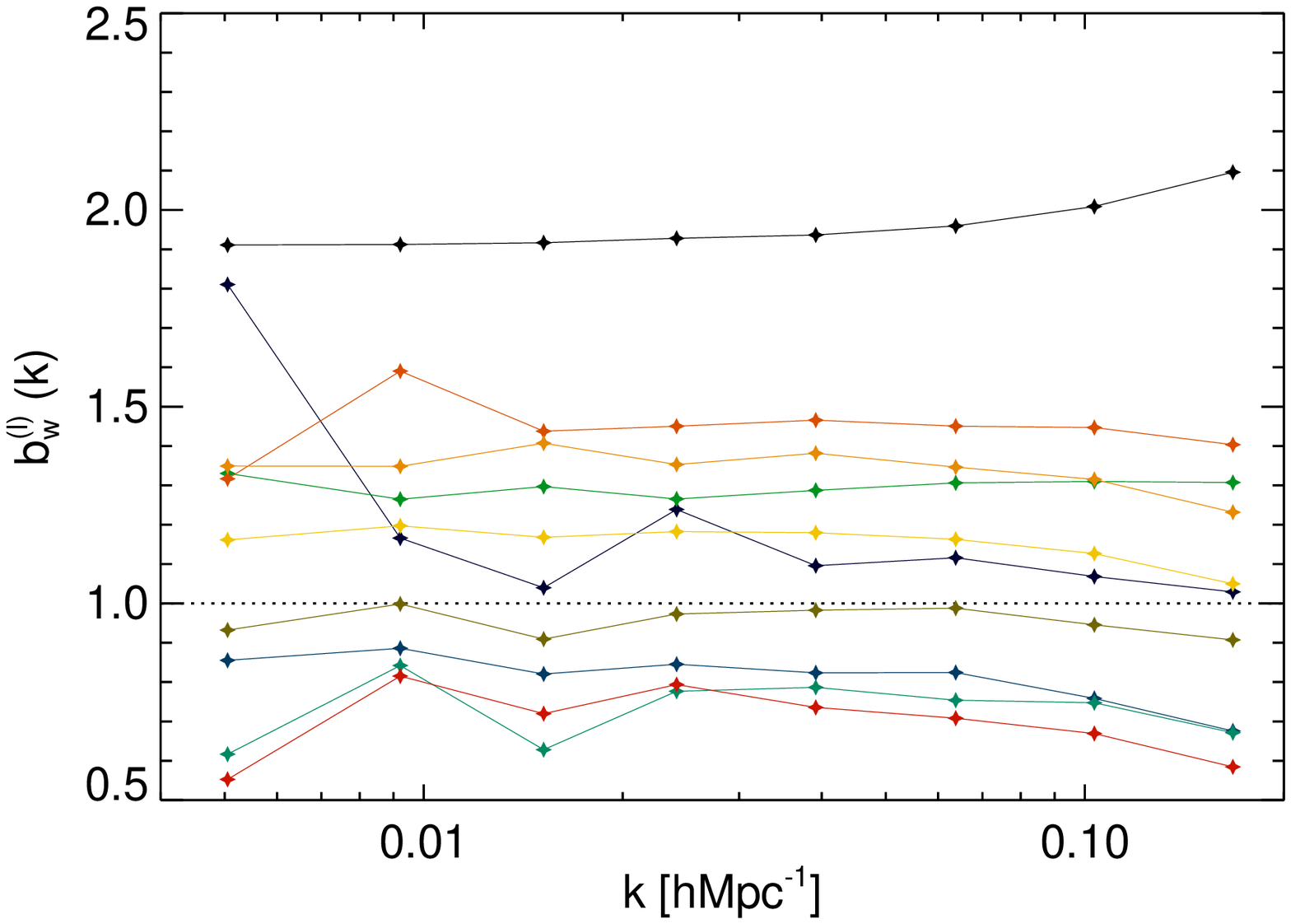}
\qquad
\includegraphics[viewport=1cm 0.4cm 19cm 13.1cm,clip]{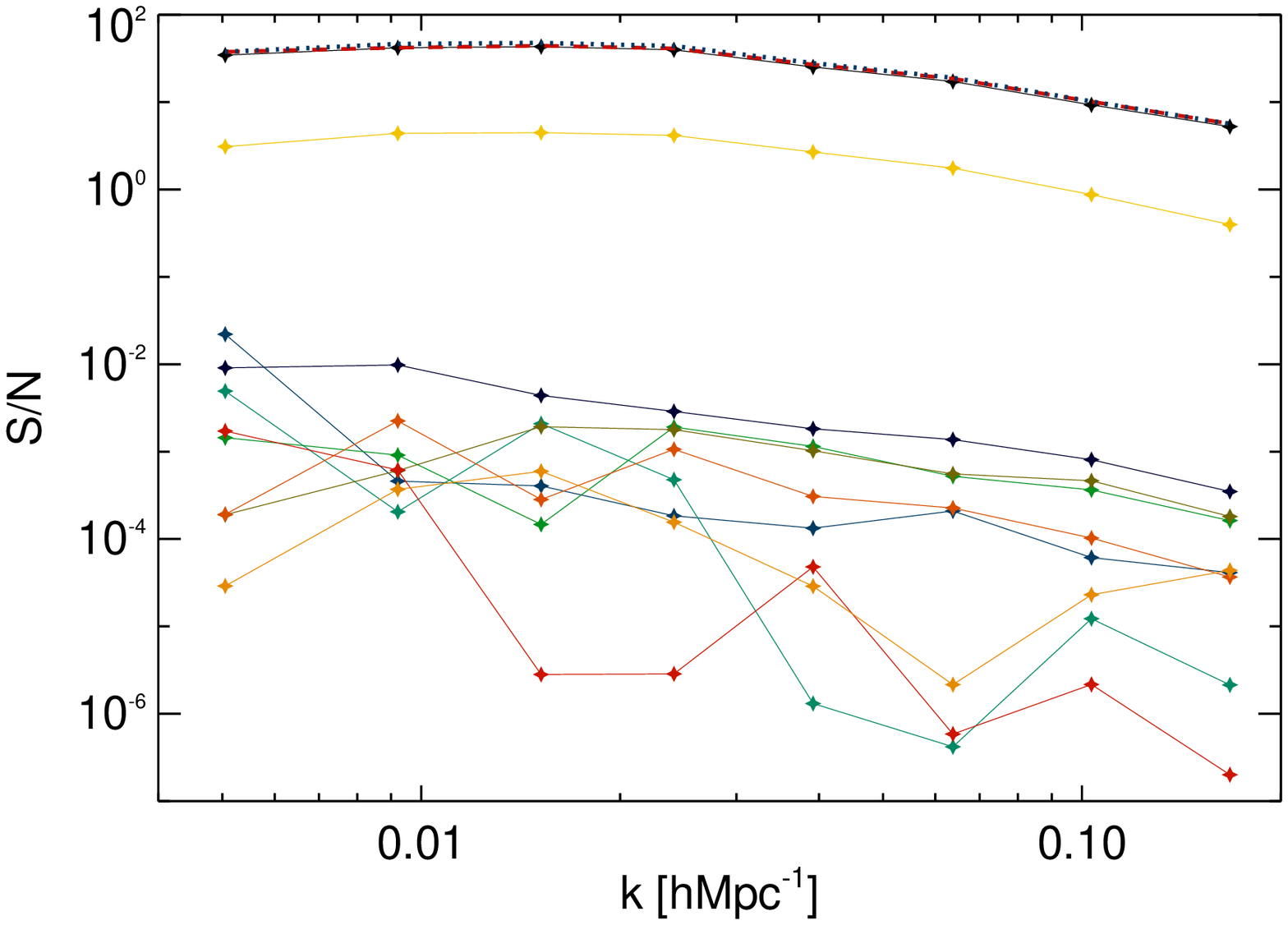}}
\caption{Weighted bias (left) and signal-to-noise ratios (right) of the 10 $V^{(l)}$-weighted halo density fields. Colors correspond to the eigenvalues and eigenvectors of Fig.~\ref{sn_ev}. The sum of all 10 signal-to-noise ratios is plotted as a dotted (blue) curve. The dashed (red) curve shows the signal-to-noise ratio as defined in Eq.~(\ref{S_N_chi^2}).}
\label{S_N}
\end{figure*}

We first show the $V_i^{(l)}$-weighted bias for the 10 halo bins in the left panel of Fig.~\ref{S_N}. The highest bias, $b_w~\simeq~2$, is achieved by weighting with $V_i^{-}$, as expected, since its components are all positive and give the largest weight to the highest halo masses. All the other eigenvectors produce lower values of the bias, distributed around unity. Note, however, that the weighted bias alone is not sufficient to describe the complete information content of the weighted halo density field. It is given by the signal-to-noise ratio in Eq.~(\ref{s_n}), which contains the weighted bias, the eigenvalue and the sums over the eigenvector components.

The total signal-to-noise ratio of the $10$ eigenmodes is shown in the right panel of Fig.~\ref{S_N}. This panel also shows the sum of all signal-to-noise ratios of each eigenmode, i.e. Eq.~(\ref{sum_s_n}), and the signal-to-noise ratio as defined in Eq.~(\ref{S_N_chi^2}) as a cross-check. Clearly, the weighting with $V_i^{-}$ dominates the signal-to-noise ratio. The eigenvector corresponding to the largest eigenvalue yields the second largest contribution, which may appear surprising since its effective bias is around unity and its eigenvalue is by far the largest. However, the sum of this eigenvector's components is large, which makes its signal-to-noise ratio dominant in comparison to the other eigenmodes. Still, it is suppressed by roughly $1$ order of magnitude compared to the weighting with $V_i^{-}$ and can be safely ignored. This fact can be cross-checked when we compare the vector $\sum_{i}C_{ij}^{-1}b_i$ appearing in Eq.~(\ref{mass_estimator}) to $V_i^{-}$. We found no mentionable discrepancy between the two. Thus, the main conclusion from this analysis is that the lowest eigenvalue contains most of the information and the other eigenmodes can be neglected.

\begin{figure*}[!t]
\centering
\resizebox{\hsize}{!}{
\includegraphics[viewport=1cm 0.4cm 19cm 13.1cm,clip]{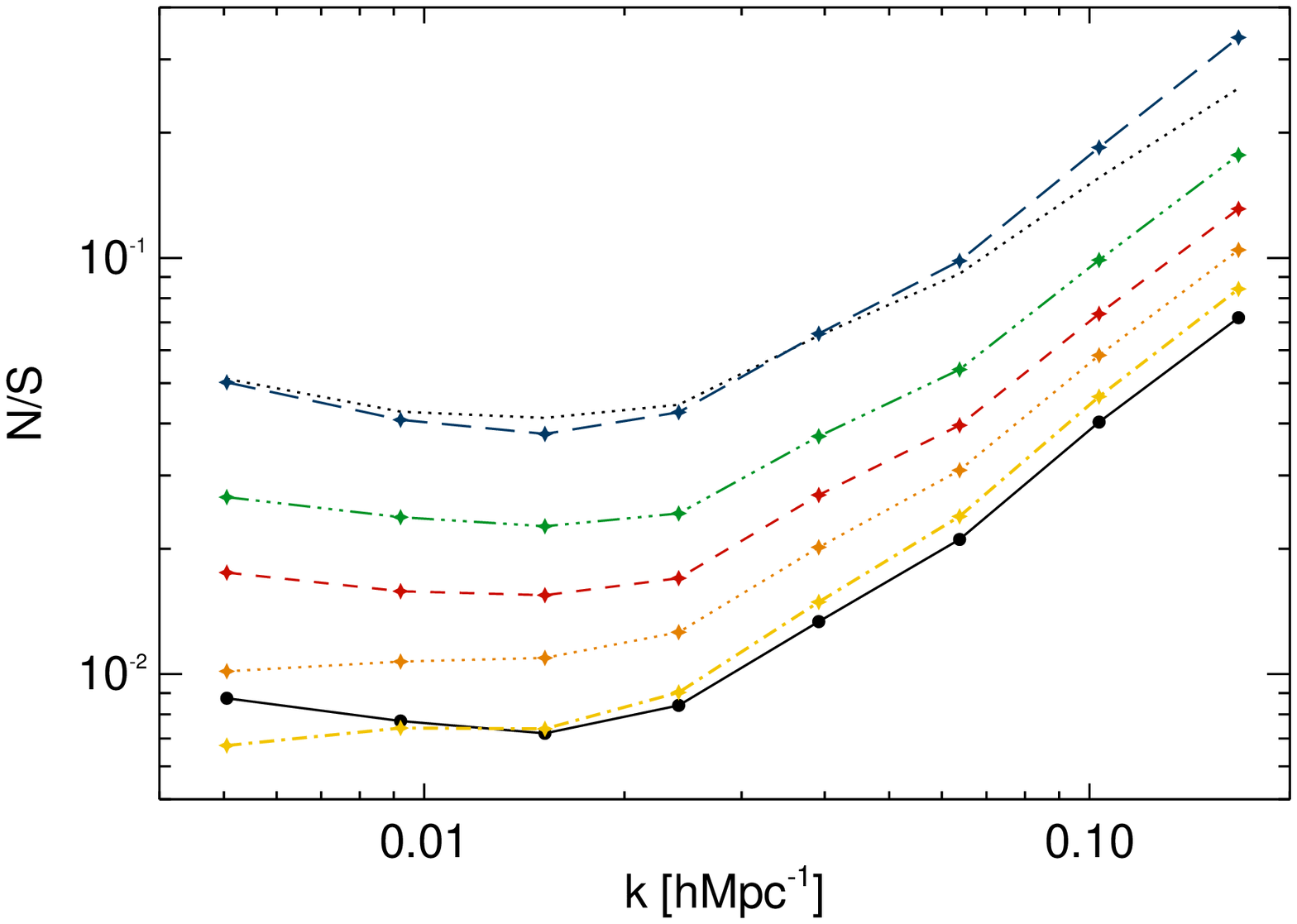}
\qquad
\includegraphics[viewport=0.6cm 0.4cm 19cm 13.1cm,clip]{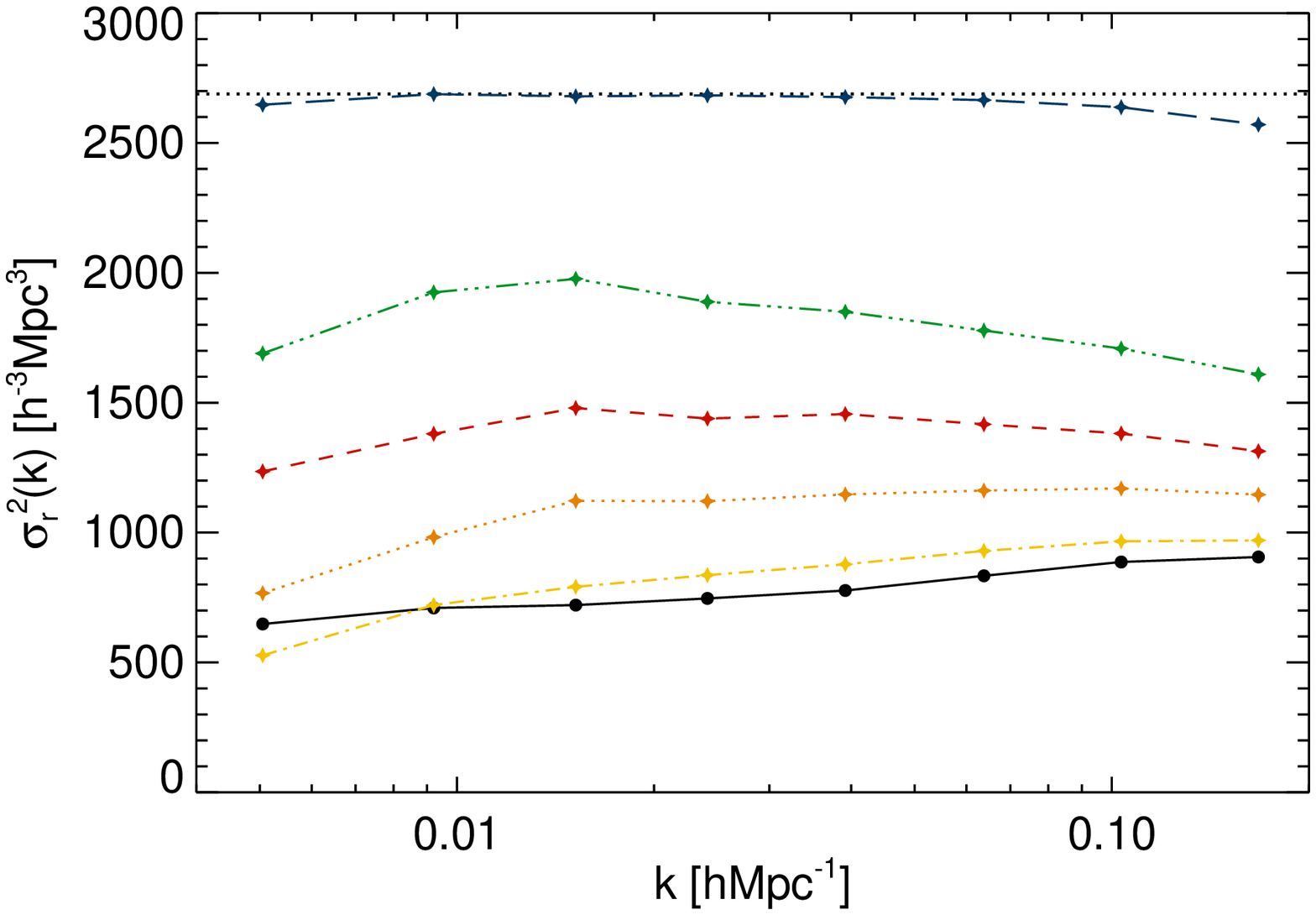}}
\caption{Inverse signal-to-noise ratio (left) and reduced shot noise (right) of the halo density field, sliced into 10 (dot-dot-dashed green line), 30 (dashed red line), 100 uniformly weighted (dotted orange line) and 100 mass-weighted (dot-dashed yellow line) mass bins. The upper curves (long dashed blue line) show the results when neglecting the off-diagonal elements of the shot noise matrix. They agree well with uniform weighting (dotted black line, left panel) and the value $1/\bar{n}$ (dotted black line, right panel), respectively. The lowest curves (solid black line) display the results obtained from weighting the halo density field with the smooth function $w(M)=M+M_0$.}
\label{sigma}
\end{figure*}

So far we explored the signal-to-noise ratio of only 10 eigenmodes, a relatively sparse mass binning of the halo density field. Do these results converge with increasing the number of halo mass bins? It is interesting to plot the inverse signal-to-noise ratio, since it appears in Eq.~(\ref{sigma_P}) and thus determines the relative error on the power spectrum. We display it in the left panel of Fig.~\ref{sigma}, where the results for different numbers of halo bins are presented. Increasing the number of halo bins improves the signal-to-noise ratio. In the limit of a large number of bins this should be equivalent to applying the smooth weighting function we found from the eigenvector $V_i^{-}$ (modified mass weighting) to each halo individually. We thus compute the inverse signal-to-noise ratio from the \emph{smoothly} weighted halo density field, defined as
\begin{equation}
\delta_w(\mathbf{x})=\frac{\sum_i w(M_i)\delta_i(\mathbf{x})}{\sum_i w(M_i)} \;. \label{smooth}
\end{equation}
In practice, for every halo of mass $M$ in the simulation we assign a weight $w(M)$ according to the smooth weighting function of Eq.~(\ref{mod_weight}). Summing over all such weighted halo overdensities and normalizing yields the smoothly weighted halo density field $\delta_w$.

\begin{figure*}[!t]
\centering
\resizebox{\hsize}{!}{
\includegraphics[viewport=1cm 0.4cm 19cm 13.1cm,clip]{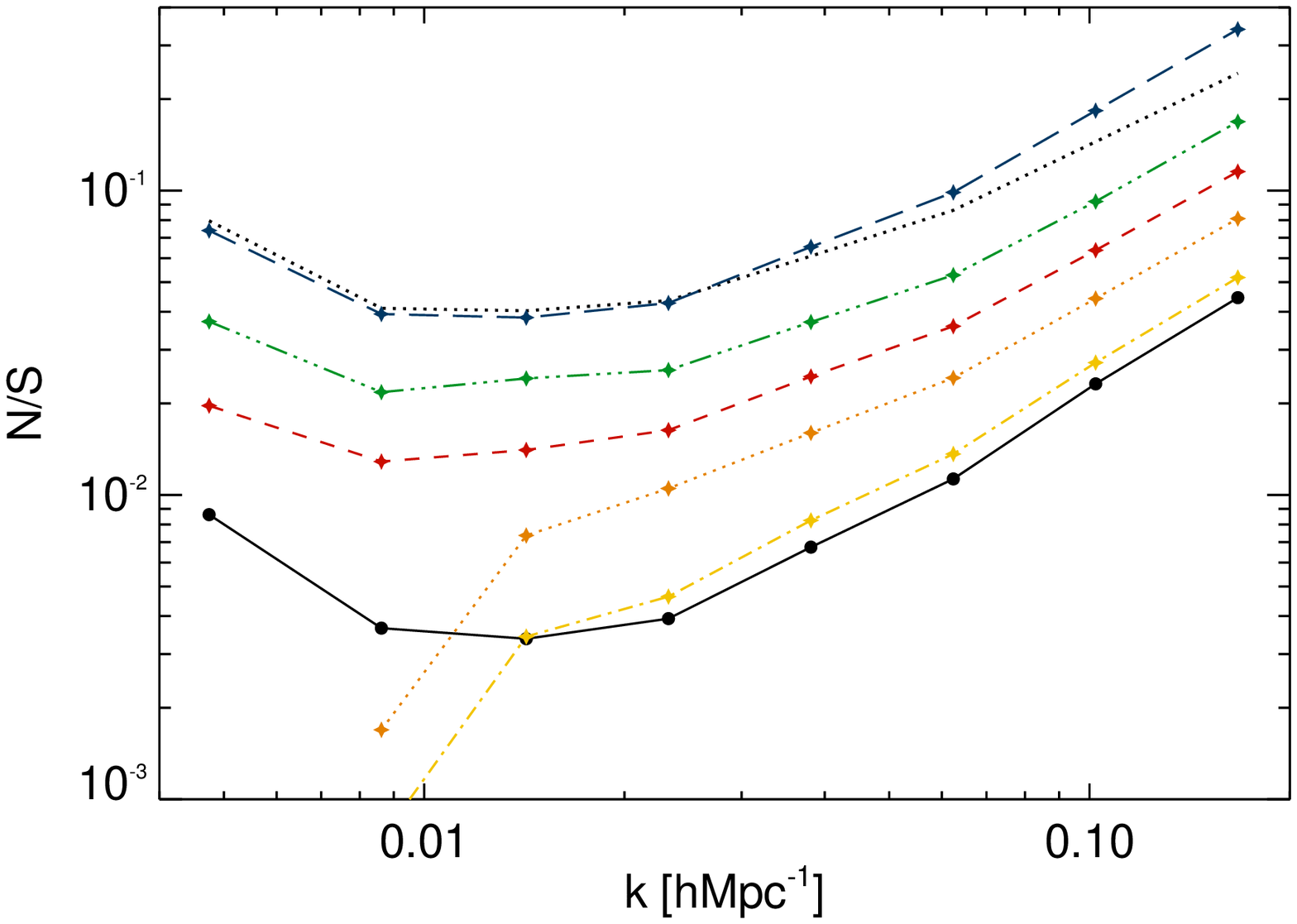}
\qquad
\includegraphics[viewport=0.6cm 0.4cm 19cm 13.1cm,clip]{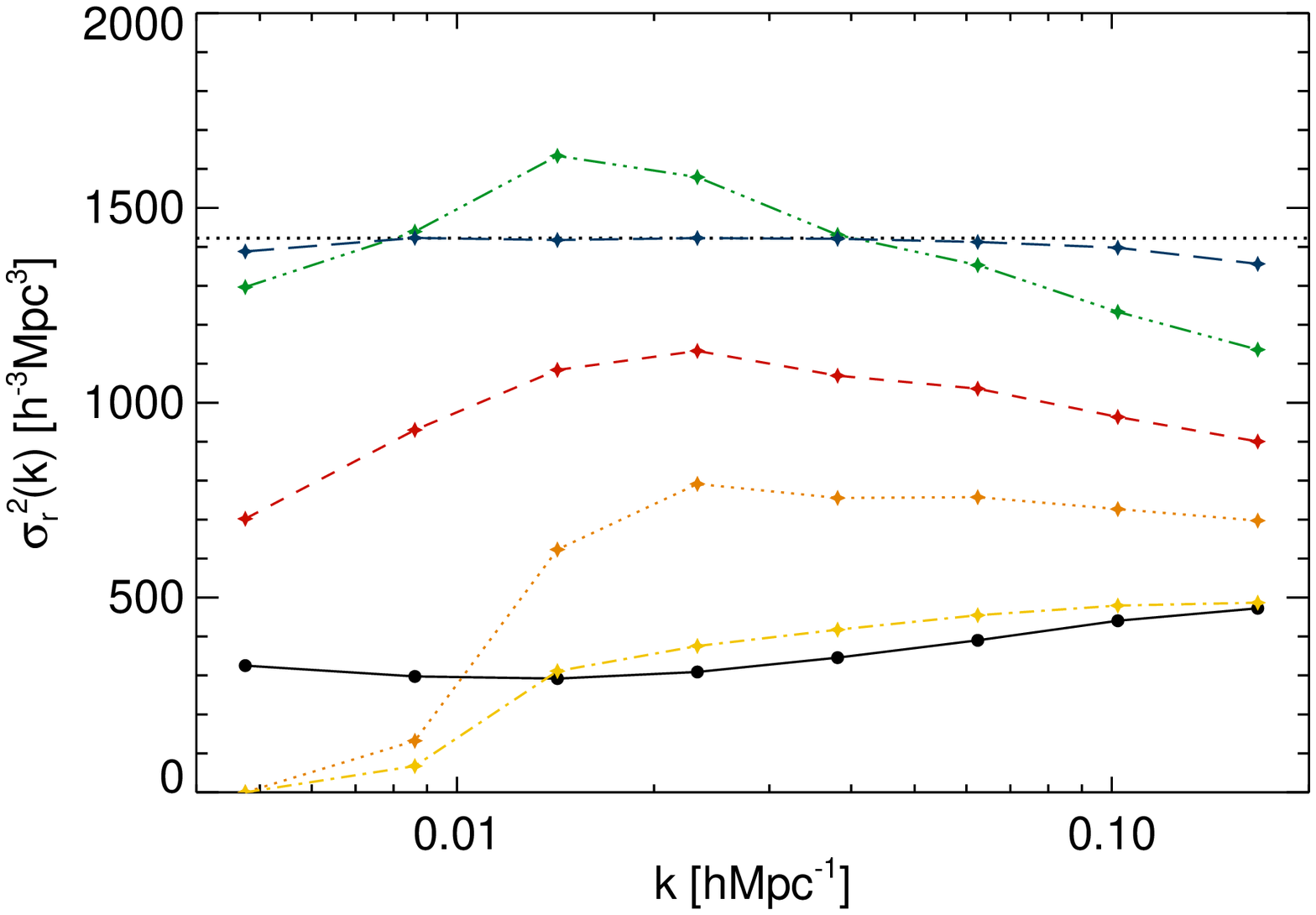}}
\caption{Same as Fig.~\ref{sigma}, computed from our higher resolution simulation with $1024^3$ particles and a mean halo number density of $\bar{n}\simeq7.0\times10^{-4}h^3\mathrm{Mpc^{-3}}$, resolving halos down to $M_{\mathrm{min}}\simeq5.9\times10^{12}h^{-1}\mathrm{M}_{\odot}$.}
\label{sigma_V}
\end{figure*}

Its inverse signal-to-noise ratio results in the lowest (solid black line) curve in the left panel of Fig.~\ref{sigma}. We also compare to the case when we assume all off-diagonal elements of the shot noise matrix to vanish (long dashed blue line). Clearly, a lot of information is lost when doing so and any improvements compared to uniform weighting (dotted black line) are canceled: we find roughly a factor of 5 improvement in the best case compared to uniform weighting. We optimized the value for $M_0$ by iteration to reach a minimal shot noise level and find $M_0\simeq3.4\times10^{13}h^{-1}\mathrm{M}_{\odot}$, larger than our best-fit values for the highest resolution eigenvector from Fig.~\ref{sn_evec1} and the vector from Fig.~\ref{opt_vector}. This is expected, since we only tested up to 100 halo mass bins and have not fully converged yet ($M_0$ increases with the number of bins). However, the results from 100 mass-weighted halo bins closely approach the results obtained from smoothly weighting the halo density field (solid black lines in Fig.~\ref{sigma}).

Looking at the reduced shot noise, we see a similar behavior. The right panel of Fig.~\ref{sigma} shows the same improvements when accounting for the off-diagonal elements of the shot noise matrix and increasing the number of bins. The shot noise of the halo density field can drastically be reduced using the appropriate weighting, on average by a factor of 4 in this case. Since the bias increases with our weighting, the improvement in the inverse signal-to-noise ratio is more striking, though.

This is well in agreement with the results in \cite{sn_letter}, where we applied linear and a different kind of modified mass weighting to halo density fields with different abundances. The modified mass-weighting function we applied there was rather a trial function that happened to suppress the shot noise better than linear mass weighting, and we did not derive it via any formal procedure like we do here. In order to directly compare to these older results, we apply modified mass weighting to one of the simulations presented in that paper. In particular, we use the simulation with $1024^3$ dark matter particles and a mean halo number density of $\bar{n}\simeq7.0\times10^{-4}h^3\mathrm{Mpc^{-3}}$, resolving halos down to $M_{\mathrm{min}}\simeq5.9\times10^{12}h^{-1}\mathrm{M}_{\odot}$. This yields the inverse signal-to-noise ratio and the reduced shot noise presented in Fig.~\ref{sigma_V}. We also show the results from the binned halo density field, as before. Note that the strong decline of the curves corresponding to 100 bins is likely due to noise at low $k$. It is the same effect present already in our first simulation, however it is pronounced here, since the number of modes per $k$-bin is lowered by a factor of 6. Compared to the best case shown in Fig.~2 of \cite{sn_letter}, we managed to further reduce the inverse signal-to-noise ratio by an additional factor of 2.

The overall improvement compared to uniform weighting is even more striking, about a factor of 10 on average in signal-to-noise and roughly a factor of 4-5 in shot noise. Hence, owing to the higher mass resolution of this simulation, we include many more low-mass halos and therefore roughly double the signal. Via iteration we find the value $M_0~\simeq~1.7\times10^{13}h^{-1}\mathrm{M}_{\odot}$ to yield the lowest shot noise level. Comparing to the value we found in the previous simulation, $M_0\simeq3.4\times10^{13}h^{-1}\mathrm{M}_{\odot}$, it is roughly half as large. In our lower resolution simulation the lowest halo mass we can resolve is $M_{\mathrm{min}}\simeq1.1\times10^{13}h^{-1}\mathrm{M}_{\odot}$, while in the higher resolution simulation it is $M_{\mathrm{min}}\simeq5.9\times10^{12}h^{-1}~\mathrm{M}_{\odot}$.

The decline of $M_0$ with the increase in the resolved halo mass fraction is expected, since one needs to account for the unresolved halos in the simulations. The relation between $M_0$ and $M_{\mathrm{min}}$ should be monotonic: in the limit of perfect mass resolution we would expect all the dark matter to be in halos of a certain mass. Weighting all these halos by their mass should then recover the statistics of the dark matter density field without shot noise, at least on large scales. Within the tested domain of our simulations we find the relation $M_0 \simeq 3M_{\mathrm{min}}$ to be a good approximation in order to determine the appropriate choice for $M_0$ given $M_{\mathrm{min}}$.

\begin{figure*}[!t]
\centering
\resizebox{\hsize}{!}{
\includegraphics[viewport=1cm 0.4cm 19cm 13.1cm,clip]{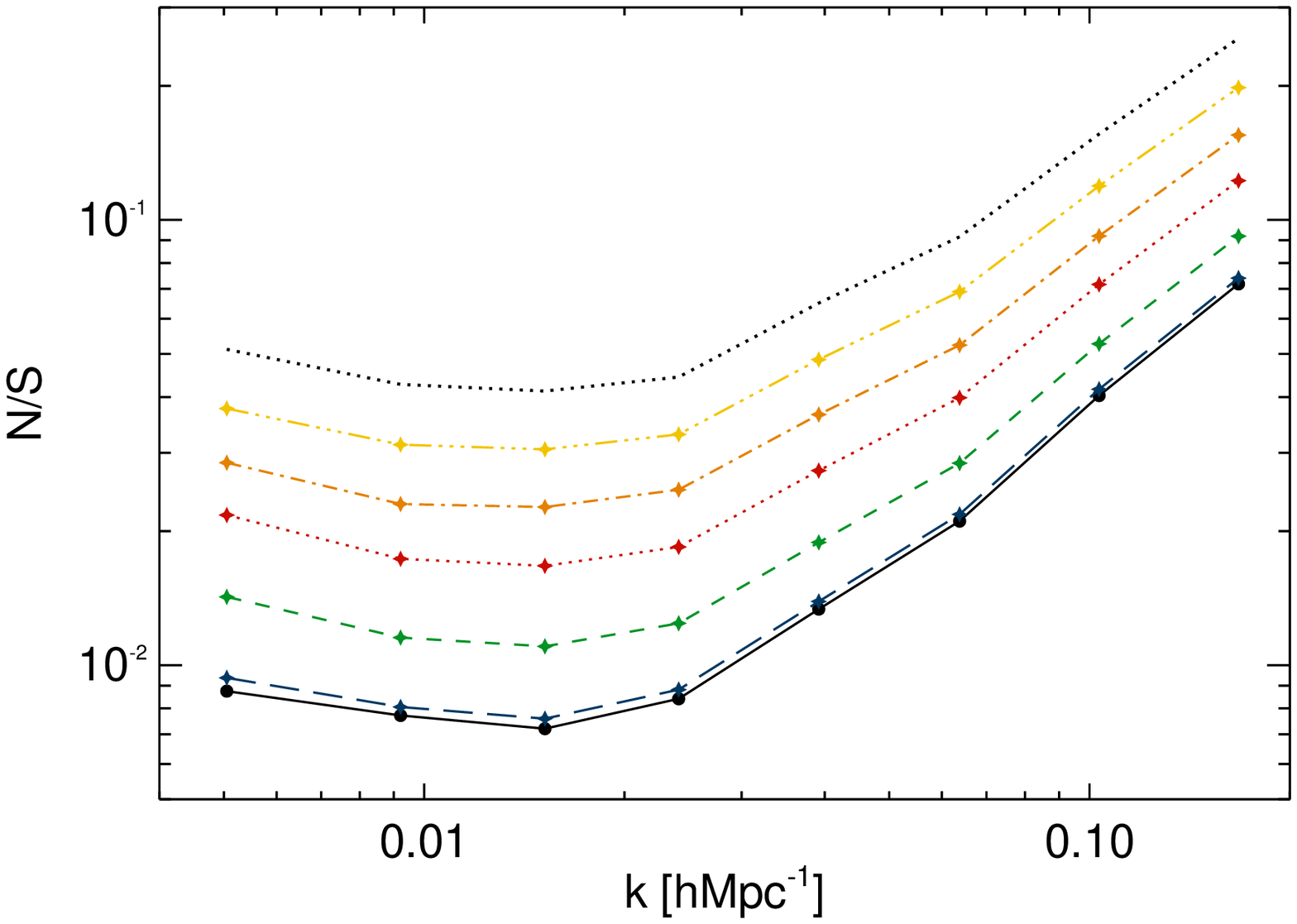}
\qquad
\includegraphics[viewport=0.6cm 0.4cm 19cm 13.1cm,clip]{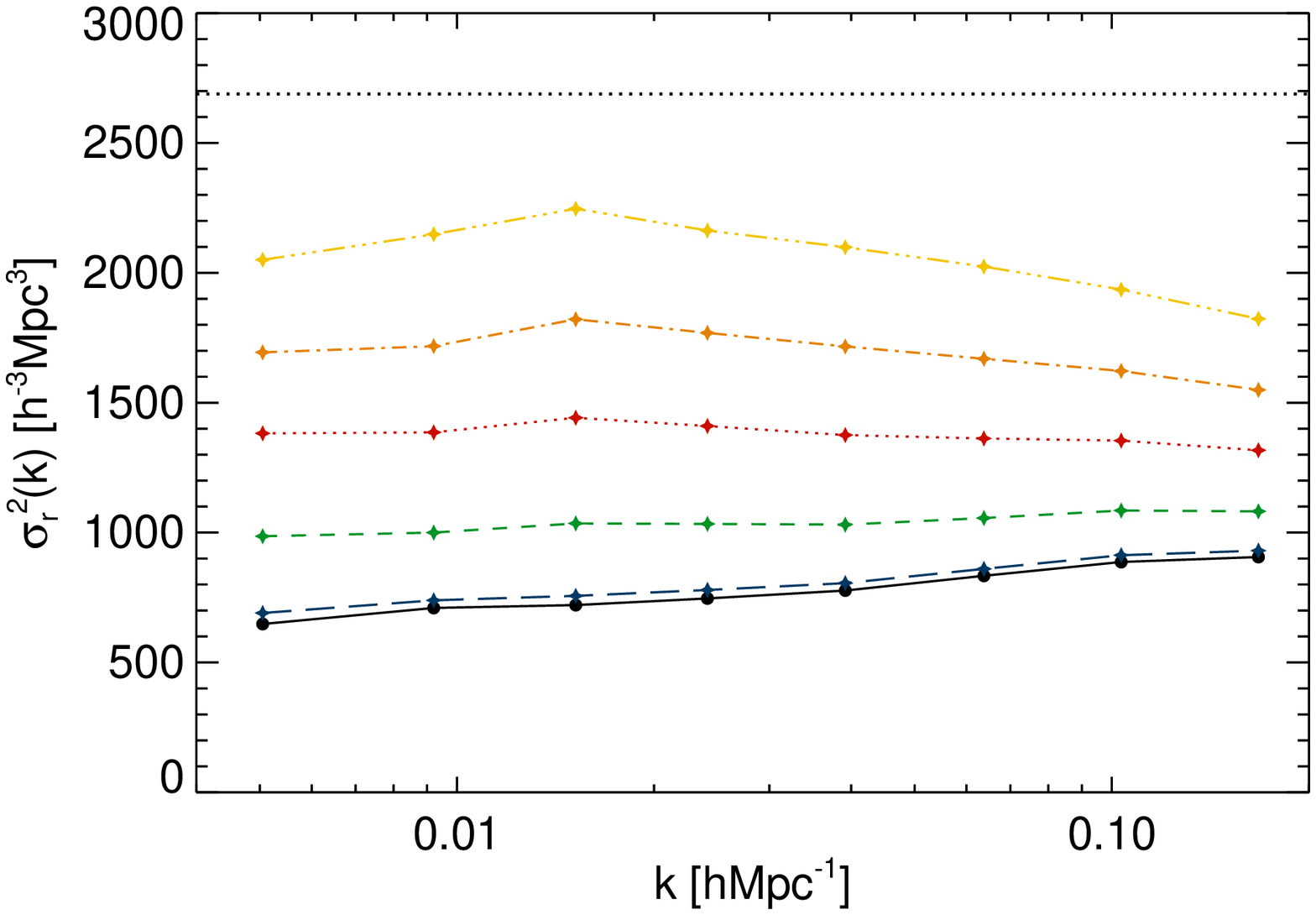}}
\caption{Inverse signal-to-noise ratio (left) and reduced shot noise (right) of halos with a log-normal mass scatter of $\sigma_{\ln M}=0,\;0.1,\;0.3,\;0.5,\;0.8\rightarrow0.4$ and $1.0$ (bottom to top), weighted by $w(M)=M+M_0$. The dotted (black) lines show the results from uniform weighting.}
\label{sigma_sc}
\end{figure*}

\subsection{Mass uncertainty}
Up to now we have always been assuming to precisely know the mass of each halo (up to the sampling variance of the halo finder). However, in realistic observations the halo mass can only be determined with a limited accuracy. Commonly, the uncertainty is expressed as a log-normal scatter in halo mass. We can mimic this uncertainty by adding a Gaussian random variable $\mathcal{G}$ with zero mean and unit variance, scaled by $\sigma_{\ln M}$, to the exponent of the mass,
\begin{equation}
\tilde{M}=M\exp(\sigma_{\ln M}\;\mathcal{G}-\sigma_{\ln M}^2/2)\; .
\end{equation}
This yields the noisier mass $\tilde{M}$, which then follows a log-normal distribution. The value $\sigma_{\ln M}$ is the log-normal scatter and the term $\sigma_{\ln M}^2/2$ is subtracted to maintain the same mean. For optical tracers of clusters $\sigma_{\ln M}$ is about $0.5$ \cite{Mass_scatter} and is expected to be much lower for SZ or X-ray proxies, such as $Y_X$ \cite{Mass_scatter2}. At the lower mass end the log-normal scatter is, however, poorly constrained. For simplicity, we will consider a constant log-normal scatter for all halos and apply the values of $\sigma_{\ln M}=0.1, \;0.3, \;0.5, \;1.0$. As a more complicated model we vary the scatter linearly with mass, with $\sigma_{\ln M}=0.8$ at $M=10^{12}h^{-1}\mathrm{M}_{\odot}$ and $\sigma_{\ln M}=0.4$ at $M=10^{15}h^{-1}\mathrm{M}_{\odot}$ (abbreviated as $\sigma_{\ln M}=0.8\rightarrow0.4$).

\begin{figure}[!t]
\centering
\resizebox{\hsize}{!}{\includegraphics[scale=0.45,viewport=1cm 0.4cm 19cm 13.1cm,clip]{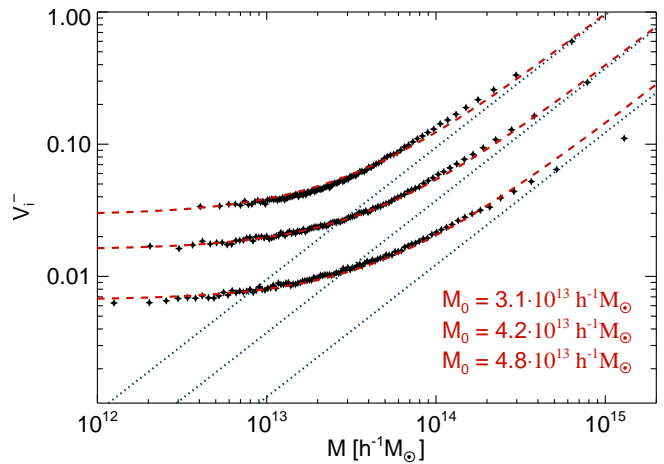}}
\caption{The normalized eigenvector $V_i^{-}$ computed for 100 uniformly weighted bins with a log-normal scatter of $\sigma_{\ln M}=0.5,\; 0.8\rightarrow0.4$ and $1.0$ added to the halo masses (top to bottom). For visibility, the lower two eigenvectors are shifted downwards by a factor of $2$ and $5$. The dotted (blue) and the dashed (red) lines represent linear and modified mass weighting, respectively.}
\label{sn_evec1_sc}
\end{figure}

We again apply modified mass weighting to construct the smoothly weighted density field as in Eq.~(\ref{smooth}) and compute the inverse signal-to-noise ratio as well as the reduced shot noise from it. For each case we adjust the value for $M_0$ in the weighting function separately by iteration. The results are depicted in Fig.~\ref{sigma_sc}. When using modified mass weighting, a $50\%$ log-normal scatter in halo mass still yields about half the shot noise level of what is expected from uniform weighting. Even our model with linearly decreasing $\sigma_{\ln M}$ from $0.8$ to $0.4$ and the high value of $\sigma_{\ln M}=1.0$ yields inverse signal-to-noise ratios and shot noise levels that are below common expectations. The optimal values for $M_0$ increase with higher mass scatter. We find $M_0\simeq3.5\times10^{13}, \;5.1\times10^{13}, \;9.0\times10^{13}, \;4.4\times10^{14}h^{-1}\mathrm{M}_{\odot}$ for the cases of $\sigma_{\ln M}=0.1, \;0.3, \;0.5, \;1.0$.

Figure~\ref{sn_evec1_sc} shows the resulting eigenvector $V_i^{-}$ when applying a log-normal scatter with $\sigma_{\ln M}=0.5,\; 0.8\rightarrow0.4$ and $1.0$ to the halo masses. In this case we only present it with uniformly weighted bins, since due to the scatter, mass weighting the bins does not improve on the results. Clearly, the saturation effect at low masses is more pronounced the stronger the scatter, resulting in an increase of the value for $M_0$. Also the halo mass range becomes wider. However, the smooth function for modified mass weighting still fits the data well. The only impact that mass scatter has on the eigenvector $V_i^{-}$ is to raise its saturation tail and thus the value of $M_0$. This is even the case for our model of linearly varying $\sigma_{\ln M}$ with mass: modified mass weighting still provides a reasonable fit to the simulation data.

\section{Halo Model approach}
In order to interpret our results, let us consider the \emph{halo model} \cite{Robert-2007,halo1,halo2,halo3,halo4,halo5,halo6,Bernardeau}. The basic assumption of this model is that the power spectra of either dark matter or halos can be written as the sum of a \emph{one-halo} term $P^{\mathrm{1H}}$ and a \emph{two-halo} term $P^{\mathrm{2H}}$. The former describes the clustering of substructure within one single halo, whereas the latter represents the clustering among different halos. Moreover, it is assumed that all the dark matter is confined within virialized halos. The two terms can be expressed analytically via the halo mass function $\frac{\mathrm{d}n}{\mathrm{d}M}(M)$, the normalized Fourier transform of the halo profile $u(k,M)$, the analytic bias $b(M)$, the linear power spectrum $P_{\mathrm{lin}}(k)$, and the mean density of dark matter $\bar{\rho}_m$. Considering all the possibilities of auto- and cross-power spectra between halos in distinct mass bins $i$ and $j$ and the dark matter, the halo model for uniform weighting yields:
\begin{widetext}
\begin{eqnarray}
&P^{\mathrm{2H}}_{ij}(k)=\frac{1}{\bar{n}_i\bar{n}_j}\iint \frac{\mathrm{d}n}{\mathrm{d}M}(M)\frac{\mathrm{d}n}{\mathrm{d}M}(M')b(M)b(M')\Theta(M,M_i)\Theta(M',M_j)P_{\mathrm{lin}}(k)\mathrm{d}M\mathrm{d}M'=b_ib_jP_{\mathrm{lin}}(k) \;& \\
&P^{\mathrm{2H}}_{im}(k)=\frac{1}{\bar{n}_i\bar{\rho}_m}\iint \frac{\mathrm{d}n}{\mathrm{d}M}(M)\frac{\mathrm{d}n}{\mathrm{d}M}(M')b(M)b(M')M'\Theta(M,M_i)P_{\mathrm{lin}}(k)\mathrm{d}M\mathrm{d}M'=b_iP_{\mathrm{lin}}(k) \;& \\
&P^{\mathrm{2H}}_{mm}(k)=\frac{1}{\bar{\rho}_m^2}\iint \frac{\mathrm{d}n}{\mathrm{d}M}(M)\frac{\mathrm{d}n}{\mathrm{d}M}(M')b(M)b(M')MM'P_{\mathrm{lin}}(k)\mathrm{d}M\mathrm{d}M'=P_{\mathrm{lin}}(k) \;& \\
&P^{\mathrm{1H}}_{ij}(k)=\frac{1}{\bar{n}_i\bar{n}_j}\int \frac{\mathrm{d}n}{\mathrm{d}M}(M)\Theta(M,M_i)\Theta(M,M_j)\mathrm{d}M=\frac{1}{\bar{n}_i}\delta^{\mathrm{K}}_{ij} \;& \label{P_hh^1h} \\
&P^{\mathrm{1H}}_{im}(k)=\frac{1}{\bar{n}_i\bar{\rho}_m}\int \frac{\mathrm{d}n}{\mathrm{d}M}(M)M\Theta(M,M_i)\mathrm{d}M=\frac{M_i}{\bar{\rho}_m} \;& \\
&P^{\mathrm{1H}}_{mm}(k)=\frac{1}{\bar{\rho}_m^2}\int \frac{\mathrm{d}n}{\mathrm{d}M}(M)M^2\mathrm{d}M\equiv\frac{\langle nM^2\rangle}{\bar{\rho}_m^2} \;.&
\end{eqnarray}
\end{widetext}
Here we assume the large-scale limit for the halo profile, i.e., $u(k\rightarrow0,M)=1$. Moreover, $\delta^{\mathrm{K}}_{ij}$ denotes the Kronecker symbol and $\Theta$ a product of two Heaviside step functions $\vartheta$:
\begin{equation}
\Theta(M,M_i)\equiv\vartheta(M-M_i)\vartheta(M_{i+1}-M) \;.
\end{equation}
Since the integrals all go from $0$ to $\infty$, this function selects the considered halo bin $i$ with mass range $M_i<M<M_{i+1}$. The corresponding average halo mass of that bin is simply denoted as $M_i$, whereas for an average over all halos we omit the index.

In the simple approach adopted here the halo model predicts a white noise term not only for the autopower spectrum of halos, but also for the halo-matter cross- and the matter autopower spectra. However, simulations have shown that the low-$k$ behavior of the dark matter one-halo term is incorrect: subtracting off the component correlated with the linear power spectrum [which is approximately $P_{\mathrm{lin}}(k)$ at large scales] from the simulated one indeed yields a $k^4$-scaling instead of a constant white noise in the residual power (mode-coupling power) \cite{halo5,Robert-2003}. This $k^4$-tail is a consequence of local mass and momentum conservation of the dark matter on small scales and the same conservation laws should also apply to halo-matter correlations. We defer further discussions of this point to a future publication, where we show that a proper implementation of mass conservation in the $k=0$ limit still yields similar results to those presented here.

The shot noise matrix as defined in Eq.~(\ref{sn}) can be written as
\begin{equation}
C_{ij}=\langle\delta_i\delta_j\rangle-b_i\langle\delta_j\delta_m\rangle-b_j\langle\delta_i\delta_m\rangle+b_ib_j\langle\delta_m^2\rangle \;. \label{sn_expand}
\end{equation}
Plugging in the sum of the corresponding one- and two-halo terms for each of the angled brackets, we see that the two-halo terms cancel each other and we are left with
\begin{equation}
C_{ij}=\frac{\delta^{\mathrm{K}}_{ij}}{\bar{n}_i}-b_i\frac{M_j}{\bar{\rho}_m}-b_j\frac{M_i}{\bar{\rho}_m}+b_ib_j\frac{\langle nM^2\rangle}{\bar{\rho}_m^2} \; . \label{sn_expand2}
\end{equation}

\begin{figure}[!t]
\centering
\resizebox{\hsize}{!}{\includegraphics[viewport=0.3cm 0.4cm 19cm 13.1cm,clip]{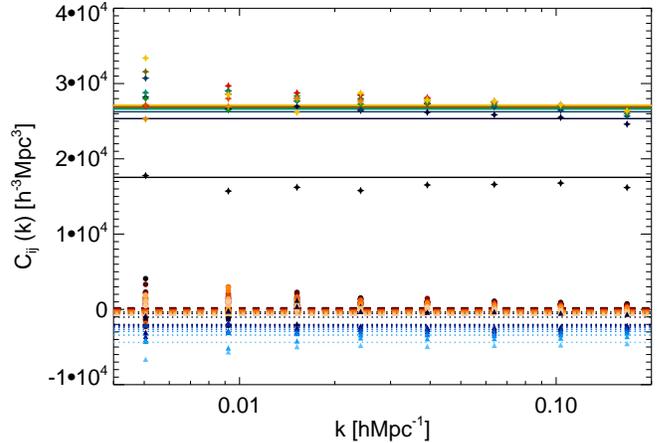}}
\caption{Elements of the shot noise matrix as described by the halo model in Eq.~(\ref{sn_expand2}), compared to the simulation results taken from Fig.~\ref{snfig} (symbols). The diagonal components (solid lines) monotonously decrease from the lowest (yellow) to the highest-mass bin (black), in good agreement with the numerical data. The halo model also reproduces both the positive (dashed lines, scaled in red) and the negative (dotted lines, scaled in blue) off-diagonal elements of the shot noise matrix fairly well.}
\label{snfig_hm}
\end{figure}

Our lower resolution simulation determines the dark matter one-halo term to be $\langle nM^2\rangle/\bar{\rho}_m^2\simeq428h^{-3}\mathrm{Mpc^3}$. Note that this value is by almost 2 orders of magnitude smaller than the first term in Eq.~(\ref{sn_expand2}). However, for highly biased halo bins it can become important in the off-diagonal terms of $C_{ij}$. The same applies to the one-halo term of the halo-matter cross-power spectrum, because it scales with the mean halo mass of each bin. For example, it yields $M_{1}/\bar{\rho}_m\simeq164h^{-3}\mathrm{Mpc^3}$ for the lowest, and $M_{10}/\bar{\rho}_m\simeq2394h^{-3}\mathrm{Mpc^3}$ for the highest of our 10 mass bins. In Fig.~\ref{snfig_hm} we compare each matrix element of Eq.~(\ref{sn_expand2}) to the numerically determined shot noise matrix (from Fig.~\ref{snfig}). The model yields a good agreement with the data, especially the observed sub-Poissonian shot noise power of the highest-mass halos, as well as the negative off-diagonal components are nicely reproduced. The off-diagonal elements with low power are more affected by scatter and therefore show stronger deviations from the theory.

For the comparison of our model to the numerical data it is, however, more convenient to look at the eigenvectors and eigenvalues of the shot noise matrix, since they describe the complete information on halo stochasticity in a more concise manner. Let us redefine the halo mass as
\begin{equation}
\mathcal{M}_i\equiv M_i-b_i\frac{\langle nM^2\rangle}{2\bar{\rho}_m} \; .
\end{equation}
Now, Eq.~(\ref{sn_expand2}) can be written more succinctly:
\begin{equation}
C_{ij}=\frac{\delta^{\mathrm{K}}_{ij}}{\bar{n}_i}-b_i\frac{\mathcal{M}_j}{\bar{\rho}_m}-b_j\frac{\mathcal{M}_i}{\bar{\rho}_m} \; .
\end{equation}

\begin{figure*}[!t]
\centering
\resizebox{\hsize}{!}{
\includegraphics[viewport=0.6cm 0.4cm 19cm 13.2cm,clip]{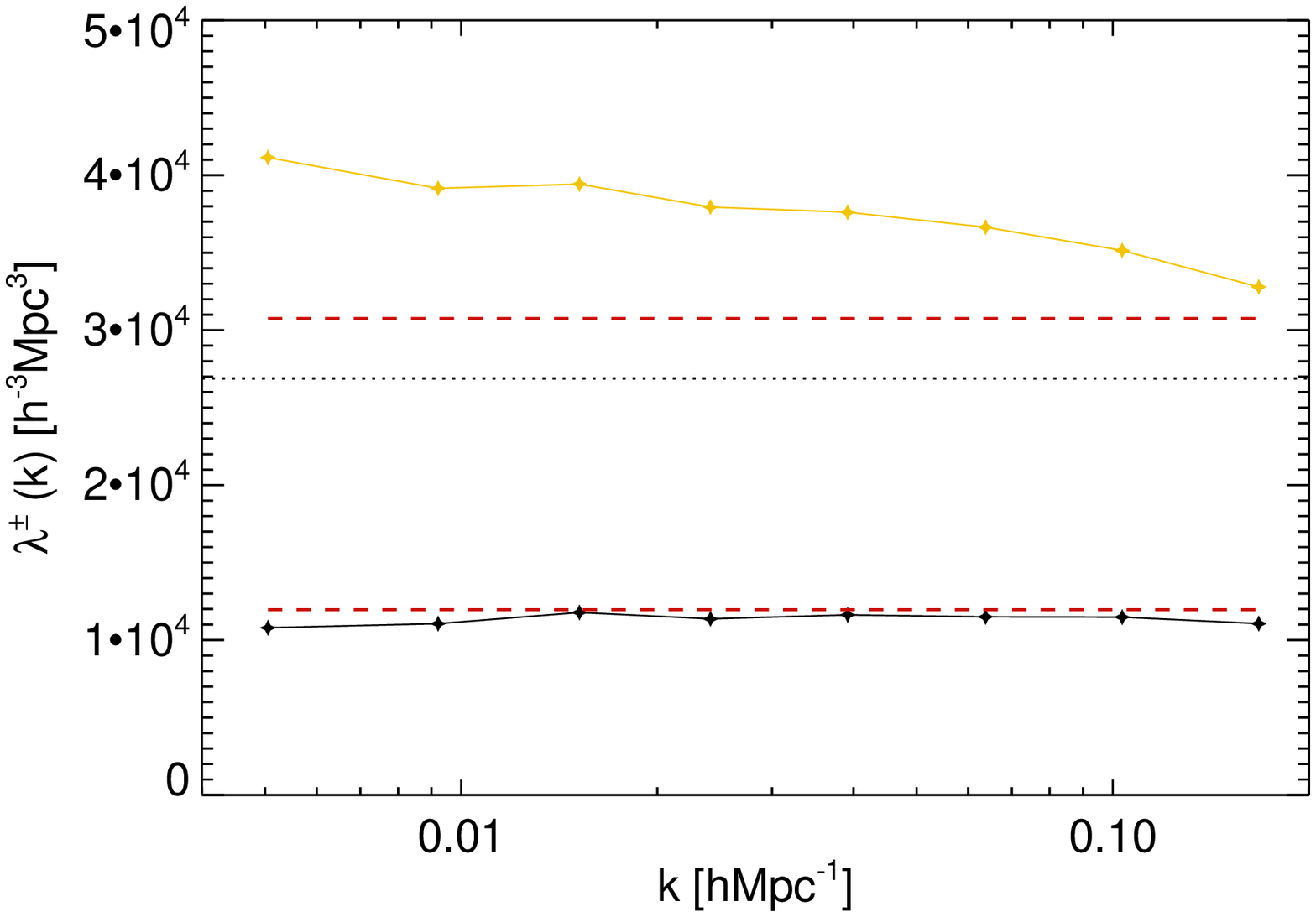}
\qquad
\includegraphics[viewport=0.4cm 0.4cm 19cm 13.1cm,clip]{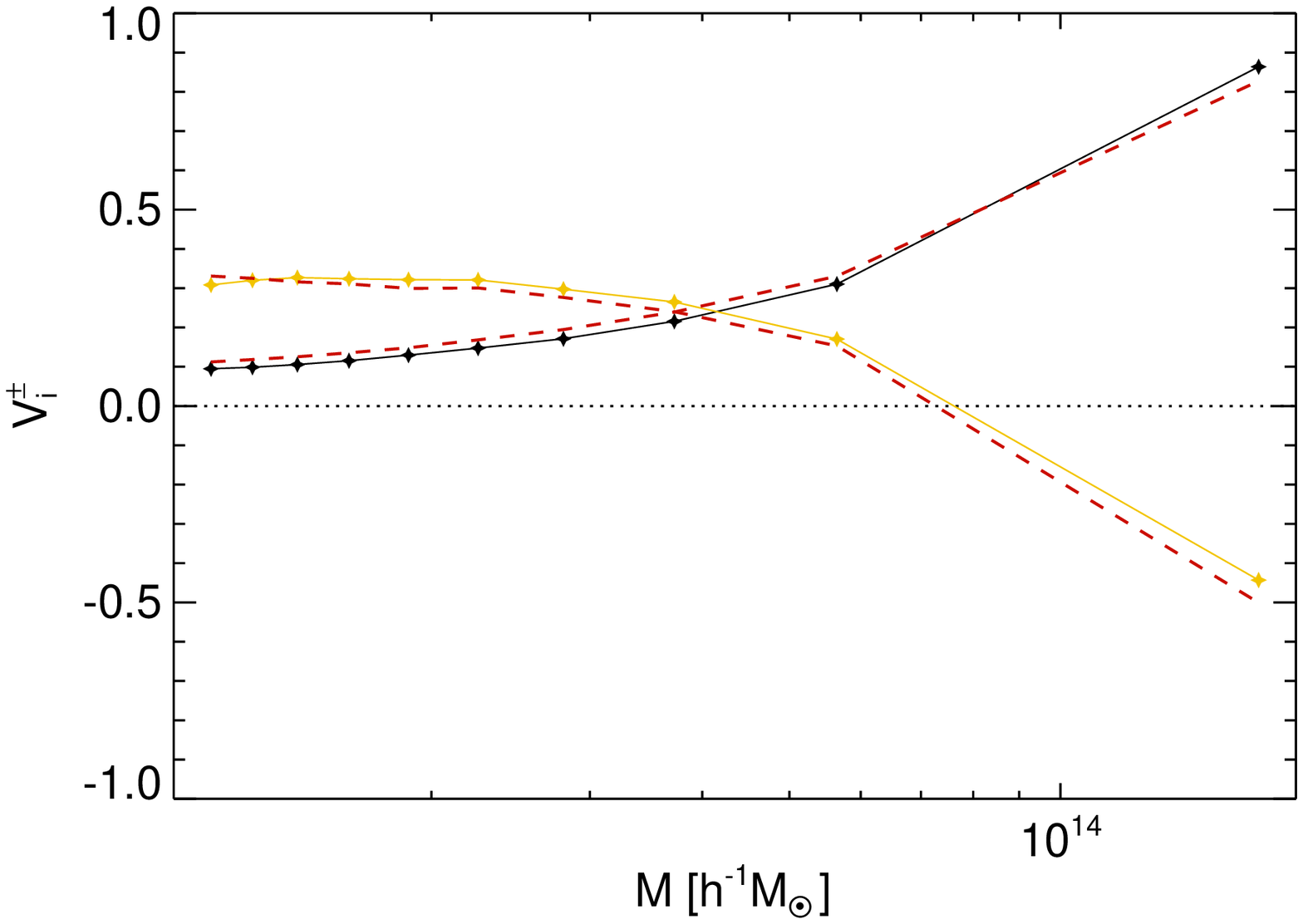}}
\caption{The eigenvalues $\lambda^{\pm}$ (left) and eigenvectors $V_i^{\pm}$ (right) from Fig.~\ref{sn_ev} compared to the predictions of the halo model (dashed red line). The dotted line in the left panel shows the value $1/\bar{n}_i$.}
\label{sn_ev_hm}
\end{figure*}

It is straightforward to work out the eigenvalues and eigenvectors of this matrix. For $d>2$ mass bins, there are $d-2$ degenerate eigenvalues with the value $\lambda=1/\bar{n}_i$. The two remaining eigenvalues with corresponding eigenvectors are
\begin{equation}
\lambda^{\pm}=\frac{1}{\bar{n}_i}-\frac{1}{\bar{\rho}_m}\sum_i\mathcal{M}_ib_i\pm\frac{1}{\bar{\rho}_m}\sqrt{\sum_i\mathcal{M}_i^2\sum_ib_i^2} \label{lambda_hm} \;,
\end{equation}
\begin{equation}
V^{\pm}_i=\frac{\mathcal{M}_i}{\sqrt{\sum_i\mathcal{M}_i^2}}\mp\frac{b_i}{\sqrt{\sum_ib_i^2}}\;. \label{V_hm}
\end{equation}
They are shown in Fig.~\ref{sn_ev_hm} for the case of 10 halo bins. It is remarkable how well the halo model reproduces the distribution of eigenvalues we found in our numerical analysis. The mass dependence of the eigenvectors also shows a good agreement. This can be seen when we renormalize $V^{\pm}_i$ by multiplication with $\sqrt{\sum_i\mathcal{M}_i^2}$ in Eq.~(\ref{V_hm}), we get
\begin{equation}
V^{\pm}_i=M_i \mp b_i\tilde{M}_0^{\pm} \;, \label{w(M)_hm}
\end{equation}
with
\begin{equation}
\tilde{M}_0^{\pm}\equiv\frac{\sqrt{\sum_i\mathcal{M}_i^2}}{\sqrt{\sum_ib_i^2}}\pm\frac{\langle nM^2\rangle}{2\bar{\rho}_m} \;. \label{M0_tilde}
\end{equation}
In other words, $V^{\pm}_i$ is nothing else than a superposition of mass and bias weighting. The relative weight between the two is determined by $\tilde{M}_0^{\pm}$. Equation~(\ref{w(M)_hm}) has a very similar form as the modified mass-weighting fitting function from Eq.~(\ref{mod_weight}). Evaluating Eq.~(\ref{M0_tilde}) using $b_i$ and $M_i$ from the simulation with 10 mass bins yields $\tilde{M}_0^-\simeq1.2\times10^{13}h^{-1}\mathrm{M}_{\odot}$. At the high-mass end, i.e., $M\simeq10^{15}h^{-1}\mathrm{M}_{\odot}$, the second term in Eq.~(\ref{w(M)_hm}) is negligible compared to the first one, since $b_i\lesssim10$ in this regime. However, at lower masses the two terms become closer in magnitude and finally the second term dominates at the low-mass end, i.e., $M\simeq10^{13}h^{-1}\mathrm{M}_{\odot}$. In this regime the bias is a slowly varying function of mass and thus well approximated by a constant. Hence, the analytical form of $V_i^{-}$ predicted by the halo model agrees well with the functional form for modified mass weighting that we found earlier.

In order to check our model more quantitatively, we compare its predictions directly to our numerical results in Fig.~\ref{sn_ev_hm}. Here we focus on the nontrivial eigenvalues $\lambda^{\pm}$ and eigenvectors $V_i^{\pm}$, since only they contain information on the halo statistics. The agreement between simulation and theory is remarkable, only for the eigenvalue $\lambda^{+}$ we find a stronger discrepancy, but since it shows a slight scale dependence it probably involves more detailed modeling. We did the same comparison for the case of 30 and 100 mass bins and find the agreement in the eigenvectors to become even better. The offset in $\lambda^{+}$ however does not vanish with an increasing number of bins.

One might argue that the way we estimate the bias from the simulation in Eq.~(\ref{bias}) is not correct in this approach, since the halo model predicts a nonzero white noise term for both the halo-matter cross, as well as the dark matter autopower spectrum. We repeated the same analysis with a shot noise corrected halo bias defined as
\begin{equation}
b_i=\frac{\langle\delta_i\delta_m\rangle-\frac{M_i}{\bar{\rho}_m}}{\langle\delta_m^2\rangle-\frac{\langle nM^2\rangle}{\bar{\rho}_m^2}} \;. \label{bias_sn}
\end{equation}
However, we find essentially no differences in the shot noise matrix and its eigenvalues and eigenvectors. As can be seen in Eq.~(\ref{sn_expand}), this is because small changes in the bias are compensated by terms of opposite sign. For the same reason it does not matter much whether we use the scale-dependent or scale-independent bias in our analysis.

\begin{figure*}[!t]
\centering
\resizebox{\hsize}{!}{
\includegraphics[viewport=0.6cm 0.4cm 19cm 13.2cm,clip]{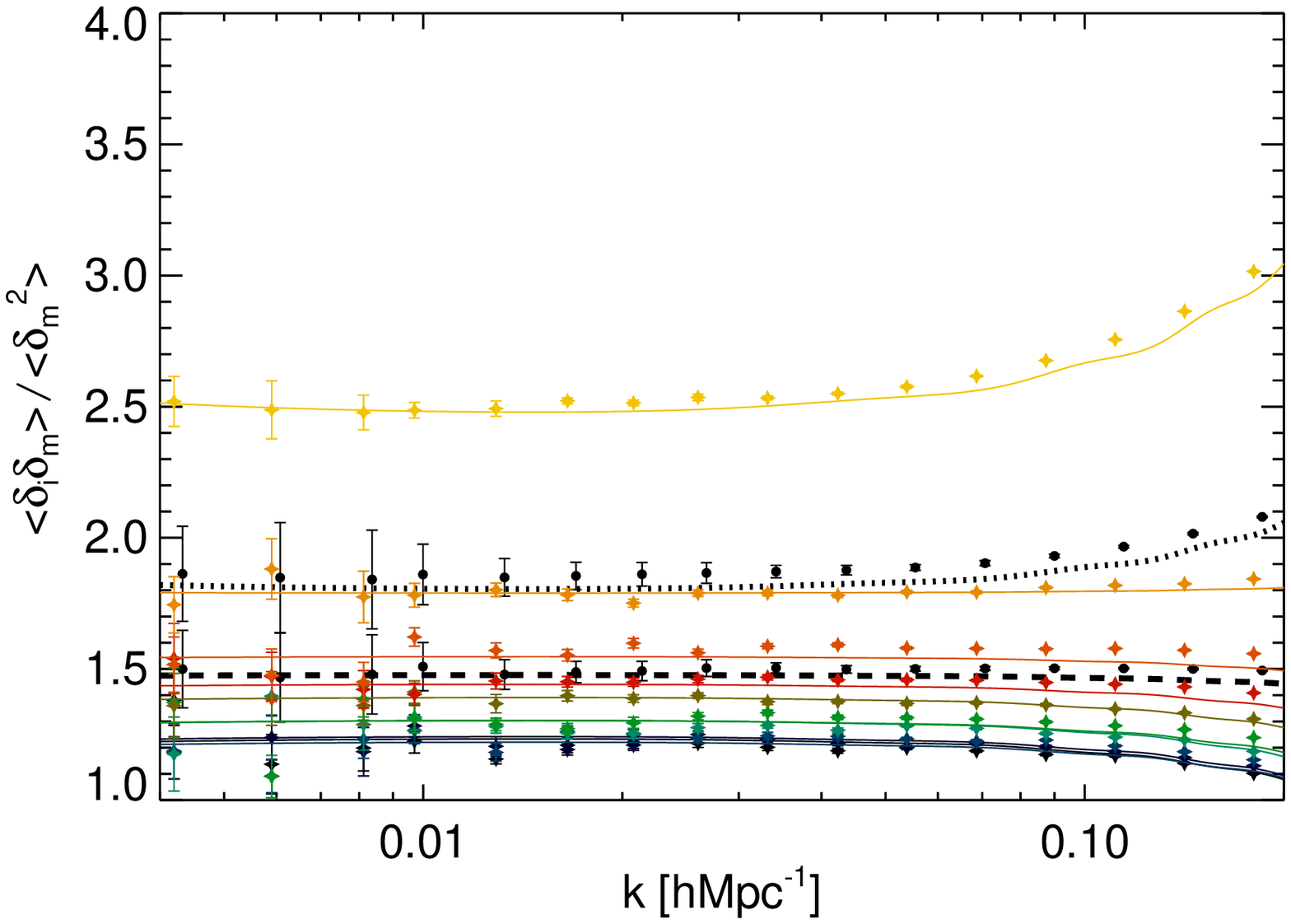}
\qquad
\includegraphics[viewport=1cm 0.4cm 19cm 13.1cm,clip]{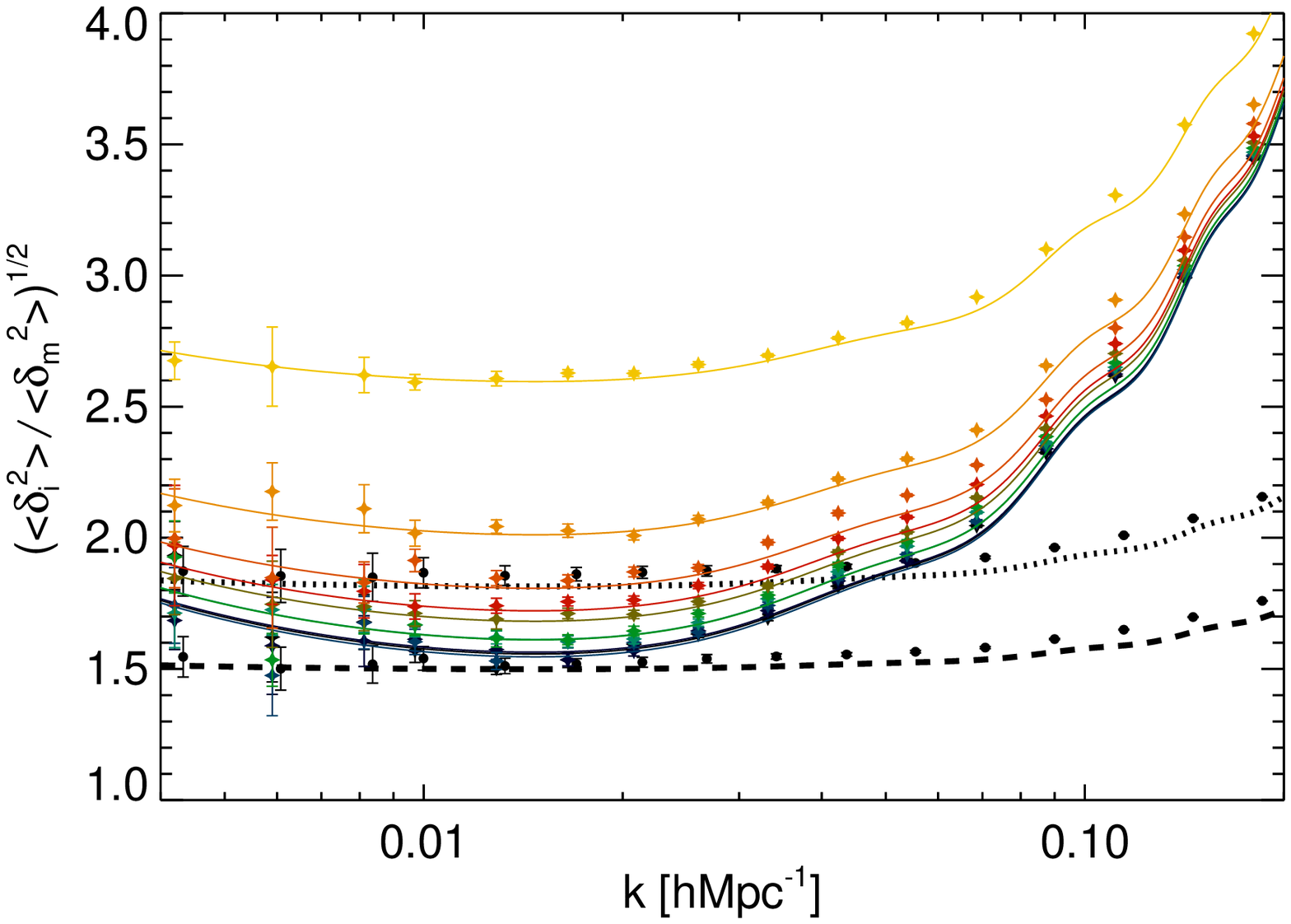}}
\caption{Scale-dependent bias estimators from halo-matter cross correlation (left panel) and halo-auto correlation (right panel) as predicted by the halo model. The simulation results from 10 mass bins are shown as crosses with error bars (in color); the solid lines show the halo model results. The black dots with error bars show the results for only one mass bin (all halos) for both uniform and modified mass weighting, with the halo model prediction overplotted in dashed and dotted, respectively.}
\label{bias_hm}
\end{figure*}

Another way to compare our model to the simulations is to look at the estimators for the halo bias itself,
\begin{equation}
\frac{\langle\delta_i\delta_m\rangle}{\langle\delta_m^2\rangle} \quad\mathrm{and}\quad \sqrt{\frac{\langle\delta_i^2\rangle}{\langle\delta_m^2\rangle}} \;. \label{estimators}
\end{equation}
Since the halo model yields white noise terms for all three correlators appearing in these estimators, this can partly account for their scale dependence. We get
\begin{eqnarray}
\frac{\langle\delta_i\delta_m\rangle}{\langle\delta_m^2\rangle}&=&\frac{b_iP_{\mathrm{lin}}(k)+\frac{M_i}{\bar{\rho}_m}}{P_{\mathrm{lin}}(k)+\frac{\langle nM^2\rangle}{\bar{\rho}_m^2}} \; , \label{estimator1} \\ \frac{\langle\delta_i^2\rangle}{\langle\delta_m^2\rangle}&=&\frac{b_i^2P_{\mathrm{lin}}(k)+\frac{1}{\bar{n}_i}}{P_{\mathrm{lin}}(k)+\frac{\langle nM^2\rangle}{\bar{\rho}_m^2}} \;, \label{estimator2}
\end{eqnarray}
where we take $b_i$ as an average over large scales from Eq.~(\ref{bias_sn}) or compute it from the halo mass function 
via the peak-background split formalism~\cite{Bias}. The simulation results for both estimators are shown in Fig.~\ref{bias_hm} for the case of 10 mass bins. The halo model reproduces the numerical results very well up to scales of $k\simeq0.1\;h\mathrm{Mpc}^{-1}$. The deviation on smaller scales is expected, since higher-order bias effects, the nonlinear evolution of the density field \cite{halo5,RPT}, and the detailed shape of the halo profile begin to matter \cite{halo4}.

The figure also shows the result of the two estimators when accounting for all of the halos in the simulation. In the case of uniform weighting (dashed lines) they both agree on large scales, but show a different scale dependence towards higher $k$-modes. With modified mass weighting (dotted lines) however, both estimators agree even up to smaller scales, a consequence of the small stochasticity in this estimator. Note that for a weighted field we need to account for the weights in the averaged quantities, so in Eqs.~(\ref{estimator1}) and (\ref{estimator2}) we have to exchange $M_i$ by the weighted mean halo mass $M_w$, $b_i$ by the weighted bias $b_w$, and $1/\bar{n}_i$ by $1/\bar{n}\times\langle w^2\rangle/\langle w\rangle^2$, with
\begin{eqnarray}
\bar{n}&=&\int\frac{\mathrm{d}n}{\mathrm{d}M}(M)\;\mathrm{d}M \;, \label{average1} \\
\langle w\rangle &=&\frac{1}{\bar{n}}\int \frac{\mathrm{d}n}{\mathrm{d}M}(M)w(M)\;\mathrm{d}M \;, \\
\langle w^2\rangle &=&\frac{1}{\bar{n}}\int \frac{\mathrm{d}n}{\mathrm{d}M}(M)w^2(M)\;\mathrm{d}M \;, \\
M_w&=&\frac{1}{\bar{n}}\int \frac{\mathrm{d}n}{\mathrm{d}M}(M)w(M)M\;\mathrm{d}M \;, \label{average2} \\
b_w&=&\frac{1}{\bar{n}}\int \frac{\mathrm{d}n}{\mathrm{d}M}(M)w(M)b(M)\;\mathrm{d}M \;, \label{average3}
\end{eqnarray}
where we integrate over all resolved halo masses.

Last but not least we can utilize the halo model to determine the reduced shot noise as a function of mass resolution. For this we need analytic expressions for the halo mass function $\frac{\mathrm{d}n}{\mathrm{d}M}(M)$ and the halo bias $b(M)$ to compute the eigenvalues and eigenvectors of the shot noise matrix. We use the functional forms of Sheth-Tormen \cite{Bias} with the parameters given in \cite{Mandelbaum}. For infinitesimal bins, Eqs.~(\ref{lambda_hm}) and (\ref{V_hm}) can be rewritten as
\begin{equation}
\lambda^{\pm}=\frac{1}{\mathrm{d}n}-\frac{1}{\bar{\rho}_m}\langle\mathcal{M}b\rangle\pm\frac{1}{\bar{\rho}_m}\sqrt{\langle\mathcal{M}^2\rangle\langle b^2\rangle} \;, \label{lambda_hm_int}
\end{equation}
\begin{equation}
V^{\pm}(M)=\frac{\mathcal{M}}{\sqrt{\langle\mathcal{M}^2\rangle}}\mp\frac{b}{\sqrt{\langle b^2\rangle}} \;, \label{V_hm_int}
\end{equation}
with
\begin{eqnarray}
\langle\mathcal{M}b\rangle&=&\frac{1}{\bar{n}}\int\frac{\mathrm{d}n}{\mathrm{d}M}(M)\mathcal{M}b\;\mathrm{d}M \;, \\
\langle\mathcal{M}^2\rangle&=&\frac{1}{\bar{n}}\int\frac{\mathrm{d}n}{\mathrm{d}M}(M)\mathcal{M}^2\;\mathrm{d}M \;, \\
\langle b^2\rangle&=&\frac{1}{\bar{n}}\int\frac{\mathrm{d}n}{\mathrm{d}M}(M)b^2\;\mathrm{d}M \;.
\end{eqnarray}
The integrals run from $M_{\mathrm{min}}$ to $\infty$ and we can compute the reduced shot noise $\sigma_r^2$ of the weighted halo density field for various values of $M_{\mathrm{min}}$ using Eqs.~(\ref{sigma_w}) and (\ref{sigma_r}) in their infinitesimal form:
\begin{equation}
\sigma_w^2(M_{\mathrm{min}})=\lambda\;\frac{\frac{1}{\bar{n}}\int\frac{\mathrm{d}n}{\mathrm{d}M}(M)V^2(M)\;\mathrm{d}M}{\left(\frac{1}{\bar{n}}\int \frac{\mathrm{d}n}{\mathrm{d}M}(M)V(M)\;\mathrm{d}M\right)^2} \;.
\end{equation}

\begin{figure}[!t]
\centering
\resizebox{\hsize}{!}{\includegraphics[scale=0.45,viewport=1cm 0.4cm 19cm 13.2cm,clip]{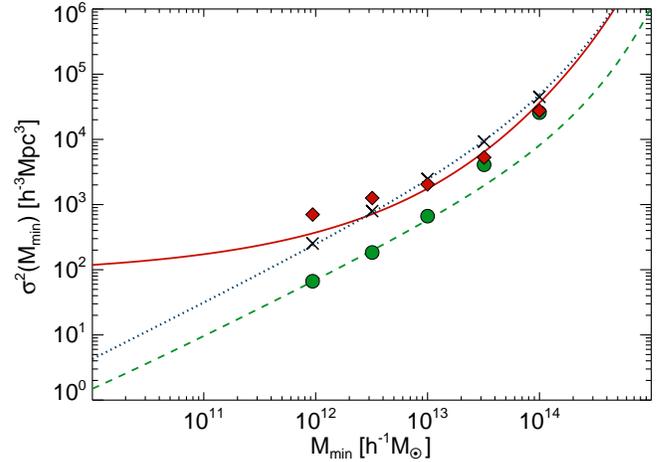}}
\caption{Stochasticity between halos and the dark matter as a function of mass resolution as predicted by the halo model for the cases of uniform ($\sigma_u^2$, solid red line) and modified mass weighting ($\sigma_r^2$, dashed green line). The dotted (blue) line shows the Poisson prediction $1/\bar{n}$. The results from our highest resolution simulation are overplotted as red diamonds (uniform weighting) and green circles (modified mass weighting) for five different low-mass cuts. The black crosses show the corresponding values of $1/\bar{n}$ taken from the simulation.}
\label{sn_hm}
\end{figure}

We neglect all eigenmodes except $\lambda^{\pm}$ and $V^{\pm}$ for this calculation, since they have the largest contribution in signal-to-noise. This yields
\begin{equation}
\sigma_r^2(M_{\mathrm{min}})=\left(\frac{1}{\sigma_{w^+}^2}+\frac{1}{\sigma_{w^-}^2}\right)^{-1} \;.
\end{equation}
The result can then be compared to the expected shot noise from uniform weighting, which, according to the halo model, is given by
\begin{equation}
\sigma_u^2(M_{\mathrm{min}})=\frac{1}{\bar{n}}-2b\frac{M}{\bar{\rho}_m}+b^2\frac{\langle nM^2\rangle}{\bar{\rho}_m^2} \;,
\end{equation}
where $\bar{n}$, $M$ and $b$ depend on $M_{\mathrm{min}}$ and can be computed from Eqs.~(\ref{average1}), (\ref{average2}) and (\ref{average3}) using uniform weights, i.e. $w(M)=1$. The functions $\sigma_u^2(M_{\mathrm{min}})$ and $\sigma_r^2(M_{\mathrm{min}})$ are depicted in Fig.~\ref{sn_hm}. Apparently, at low resolution (high $M_{\mathrm{min}}$), the improvement due to modified mass weighting is quite modest. However, for $M_{\mathrm{min}}\lesssim10^{12}h^{-1}\mathrm{M}_{\odot}$ the function $\sigma_u^2(M_{\mathrm{min}})$ approaches a constant, while $\sigma_r^2(M_{\mathrm{min}})$ still decreases linearly with $M_{\mathrm{min}}$. This linear trend leads to a suppression of stochasticity by almost 2 orders of magnitude if one can resolve halos down to $M_{\mathrm{min}}=10^{10}h^{-1}\mathrm{M}_{\odot}$.

In order to cross-check these results we computed the $k$-averaged shot noise (shown as filled symbols) for various low-mass cuts from our highest resolution simulation consisting of $1536^3$ particles resolving halos down to $M_{\mathrm{min}}\simeq9.4\times10^{11}h^{-1}\mathrm{M}_{\odot}$. Taking into account all halos in this simulation we obtain a minimal shot noise level when applying modified mass weighting with $M_0 \simeq 3.1 \times 10^{12}h^{-1}\mathrm{M}_{\odot}$, which again satisfies the anticipated relation $M_0\simeq3M_{\mathrm{min}}$.

Overall, the agreement between the simulations and the halo model is reasonable, but not perfect. At low $M_{\mathrm{min}}$ the halo model underestimates the shot noise of the uniformly weighted halo density field and the shot noise suppression due to modified mass weighting relative to uniform weighting is even larger than predicted by the model. At high $M_{\mathrm{min}}$ the shot noise in the simulation is not perfectly scale independent anymore and since we are taking the average over the whole $k$-range the result becomes more inaccurate.

\section{Conclusions}
In a previous paper \cite{sn_letter} it was shown that weighting dark matter halos by their mass can lead to a suppression of stochasticity between halos and the dark matter relative to naive expectations. In this work we investigated the shot noise matrix, defined as the two-point correlator $C_{ij}\equiv\langle(\delta_i -b_i\delta_m)(\delta_j-b_j\delta_m)\rangle$ in Fourier space, split into equal number density mass bins. The eigensystem of this matrix reveals two nontrivial eigenvalues, one of them being enhanced, the other suppressed compared to the Poisson model expectation. It is the latter that leads to a reduced stochasticity. The optimal estimator of the dark matter and the eigenvector corresponding to the lowest eigenvalue are very similar and the latter dominates the signal-to-noise ratio of the halo density field. We fit both vectors by a smooth function of mass which we denote \emph{modified mass weighting}. It is proportional to halo mass at the high-mass end and approaches a constant towards lower masses which is determined by the minimum halo mass resolved in the simulations. This constant is roughly 3 times the minimum halo mass over the range of masses we explored.

Applying this function to weight the halo density field results in a field that is more correlated with the dark matter with a suppressed shot noise component, improving upon previous results \cite{sn_letter} by a factor of 2 in signal-to-noise. We investigate the effect of uncertainty in halo mass, finding that it does not change our fundamental conclusions, even if it weakens the strength of the method: a realistic amount of log-normal scatter in mass at the level of 0.5 increases the shot noise by a about a factor of 2. Our results can directly be applied to methods that attempt to eliminate sampling variance by investigating the relation between galaxies and the dark matter both tracing the same LSS. In this case the error is determined by the stochasticity between the two and reducing it can improve the ultimate reach of these methods \cite{Pen,Stochasticity}.

Considering the halo model as a theoretical approach to describe the shot noise matrix, we find analytical expressions for its eigenvalues and eigenvectors. In particular, the two nontrivial eigenvectors can be written as a linear combination of halo bias and halo mass, which yields a considerable agreement with our simulation results. Furthermore, the two estimators of the scale-dependent bias, $\langle\delta_h\delta_m\rangle/\langle\delta_m^2\rangle$ and $\sqrt{\langle\delta_h^2\rangle/\langle\delta_m^2\rangle}$, are well reproduced. However, our model suffers from the lack of mass and momentum conservation: its implementation, together with higher-order perturbation theory and halo exclusion, further improves the agreement and will be presented elsewhere.

The halo model suggests the stochasticity between modified mass-weighted halos and the dark matter to decrease linearly with mass resolution below $M\simeq10^{12}h^{-1}\mathrm{M}_{\odot}$, yielding a suppression by almost 2 orders of magnitude at $M_{\mathrm{min}}\simeq10^{10}h^{-1}\mathrm{M}_{\odot}$ as compared to uniformly weighted halos. While we focused on the question of how well halos can reconstruct the dark matter, our analysis is also applicable to the study of stochasticity between halos themselves. Indeed, reducing the stochasticity between different halo tracers by optimal weighting techniques, while at the same time canceling sampling variance, should be possible even if the dark matter field is not measured. This will be addressed in more detail in a future work.

Specific applications are the best way to test the efficiency of our method. There is probably not much advantage in applying it to the standard power spectrum determination, where the sampling variance error dominates the error budget in the limit of small stochasticity, while in the opposite limit of rare halos, when the shot noise power is comparable to the intrinsic halo power, we do not see much gain (as demonstrated by the fact that all the points in Fig.~\ref{sn_hm} overlap for the highest $M_{\mathrm{min}}$, which corresponds to halos with the lowest number density). More promising applications are those where the sampling variance error is eliminated and the error budget is dominated by stochasticity, or the ratio of shot noise power to the halo power.

In this paper we have focused on the bias determination from galaxy and dark matter correlations \cite{Pen} as a specific application, but other applications are possible, such as constraining $f_{\mathrm{NL}}$ from non-Gaussianity \cite{fnl_constraint,Slosar} and the redshift-space parameter $\beta$ from redshift-space distortions \cite{beta,beta_cv}, to name a few. Upcoming surveys like SDSS-III \cite{SDSS-III}, JDEM/EUCLID \cite{JDEM,EUCLID} or BigBOSS \cite{BigBoss} and LSST \cite{LSST} will increase the available number of galaxies significantly, providing both 3D galaxy maps and 2D to 3D dark matter maps (via weak lensing techniques, enhanced by lensing tomography).

Our results suggest that correlating modified halo mass-weighted galaxies against the dark matter has the potential to lead to dramatic improvements in the precision of cosmological parameter estimation. We will explore more explicit demonstrations of the above mentioned applications in future work.

\begin{acknowledgments}
We thank Patrick McDonald and Martin White for useful discussions, V. Springel for making public his {\scshape gadget ii} code and for providing his {\scshape b-fof} halo finder, and Roman Scoccimarro for making public his {\scshape 2lpt} initial conditions code. RES acknowledges support from a Marie Curie Reintegration Grant and the Swiss National Foundation. This work is supported by the Packard Foundation, the Swiss National Foundation under Contract No. 200021-116696/1, and WCU Grant No. R32-2009-000-10130-0.
\end{acknowledgments}

\end{document}